\documentclass[prx, twocolumn, superscriptaddress,longbibliography]{revtex4-1}
\usepackage{amsmath}
\usepackage{amssymb}
\usepackage{amsthm}
\usepackage{amsfonts}
\usepackage{listings}
\usepackage{diagbox}
\usepackage{enumerate}
\usepackage{latexsym}
\usepackage{color}
\usepackage{xcolor}
\usepackage{bm}
\usepackage{hyperref}
\usepackage{graphicx}
\usepackage[FIGTOPCAP]{subfigure}
\usepackage{physics} 
\hypersetup{
    pdfnewwindow=true, colorlinks=true,
    linkcolor=blue, anchorcolor=blue,
    citecolor=blue, filecolor=blue,
    menucolor=blue, urlcolor=blue}

\newcommand{\nc}{\newcommand}
\nc{\webirpw}{\href{https://github.com/zjwang11/ir2pw}{\texttt{IR2PW}} }
\nc{\webirtb}{\href{https://github.com/zjwang11/ir2pw}{\texttt{IR2PH}} }
\nc{\webchecktopmat}{\href{https://www.cryst.ehu.es/cryst/checktopologicalmagmat}{\texttt{Check Topological Mat}}}
\nc{\webposbr}{\href{https://github.com/zjwang11/UnconvMat/blob/master/src_pos2aBR.tar.gz}{\texttt{pos2aBR}} }
\nc{\webUnconvMat}{\href{http://tm.iphy.ac.cn/UnconvMat.html}{\texttt{UnconvMat}} }
\nc{\online}{\href{http://tm.iphy.ac.cn/UnconvMat.html}{online}}

\nc{\ii}{{\text{i}}} 
\nc{\eq}{= & \;}
\nc{\q}[1]{Eq. (\ref{#1})}
\nc{\fig}[1]{Fig. \ref{#1}}
\nc{\figs}[1]{Figs. \ref{#1}}
\nc{\app}[1]{Appendix \ref{#1}}
\nc{\tab}[1]{Table \ref{#1}}
\nc{\vG}{\vb*{G}}
\nc{\vk}{\vb*{k}}
\nc{\br}{\vb*{r}}
\nc{\ba}{\vb*{a}}
\nc{\bb}{\vb*{b}}
\nc{\bc}{\vb*{c}}
\nc{\hatn}{\hat{\vb*{n}}}
\nc{\hata}{\hat{\vb*{a}}}
\nc{\hatb}{\hat{\vb*{b}}}
\nc{\hatc}{\hat{\vb*{c}}}
\nc{\hatx}{\hat{\vb*{x}}}
\nc{\haty}{\hat{\vb*{y}}}
\nc{\hatz}{\hat{\vb*{z}}}
\nc{\blue}[1]{{\color{blue}{#1}}}
\nc{\magenta}[1]{{\color{magenta}{#1}}}

\nc{\dg}{\dagger}
\nc{\ua}{\uparrow}
\nc{\da}{\downarrow}

\nc{\ie}{i.e., }
\nc{\eg}{e.g., }
\nc{\ea}{\textit{et al}. }

\nc{\eV}{\text{ eV}}
\nc{\meV}{\text{ meV}}
\nc{\AAA}{$\:\text{\AA}$}
\nc{\trans}[1]{{#1}^{\scriptscriptstyle \text{T}}}

\nc{\beginsupplement}{

	}

\begin{document}
\tolerance 10000
\draft
\title{Large shift current, $\pi$ Zak phase and unconventional nature in Se and Te}

\author{Ruihan Zhang}
\thanks{These authors contributed equally to this work.}
\affiliation{Beijing National Laboratory for Condensed Matter Physics,
and Institute of Physics, Chinese Academy of Sciences, Beijing 100190, China}
\affiliation{University of Chinese Academy of Sciences, Beijing 100049, China}

\author{Junze Deng}
\thanks{These authors contributed equally to this work.}
\affiliation{Beijing National Laboratory for Condensed Matter Physics,
and Institute of Physics, Chinese Academy of Sciences, Beijing 100190, China}
\affiliation{University of Chinese Academy of Sciences, Beijing 100049, China}

\author{Yan Sun}
\affiliation{Shenyang National Laboratory for Materials Science, Institute of Metal Research,
 Chinese Academy of Science, 110016 Shenyang, Liaoning, China}
\affiliation{School of Materials Science and Engineering, University of Science and Technology of China, 110016 Shenyang, Liaoning, China}

\author{Zhong Fang}
\affiliation{Beijing National Laboratory for Condensed Matter Physics,
and Institute of Physics, Chinese Academy of Sciences, Beijing 100190, China}
\affiliation{University of Chinese Academy of Sciences, Beijing 100049, China}

\author{Zhaopeng Guo}
\email{zpguo@iphy.ac.cn}
\affiliation{Beijing National Laboratory for Condensed Matter Physics,
and Institute of Physics, Chinese Academy of Sciences, Beijing 100190, China}

\author{Zhijun Wang}
\email{wzj@iphy.ac.cn}
\affiliation{Beijing National Laboratory for Condensed Matter Physics,
and Institute of Physics, Chinese Academy of Sciences, Beijing 100190, China}
\affiliation{University of Chinese Academy of Sciences, Beijing 100049, China}

\begin{abstract}
Recently, unconventional materials (or obstructed atomic insulators) have attracted much attention owing to the unconventional feature of mismatch between Wannier centers and atomic positions. In this paper, we demonstrate that the trigonal selenium and tellurium host an unconventional nature in both electronic and phonon spectra. In electronic band structures, the band representation (BR) decomposition for occupied bands has to contain the essential BR of $A@3b$, and the real-space invariant is $\delta_1@3b=-1$. The $\pi$ Zak phase suggests that the one-dimensional Se/Te chain is a chiral Su-Schrieffer-Heeger chain. The effective magnetism can be induced by $p$ states at ends. More importantly, a large shift current is obtained in Se quantum well. In addtion, in phonon spectra, three sets of phonon bands are well separated and assigned to $B@3b$, $B@3a$, and $A@3b$ BRs, respectively. Thus, the obstructed phonon states are predicted on the (0001) surface. As the prototypes of unconventional materials in both electronic and phonon spectra, our findings could create much interest in the study of obstructed surface electronic and phonon states in these novel materials.
\end{abstract}
\maketitle

\paragraph*{Introduction.}
Although topological quantum chemistry (TQC) \cite{TQCBradlyn2017, MTQC2021Luis, GaojcMagUnconvMat} and symmetry-indicator classification theory \cite{SymmIndicator2017Po, MSymmIndicator2018Po, SymmIndi2018Song, PhysRevX.7.041069SymmIndicator, peng2021topological} have achieved great success in the searching for topological band structures in both nonmagnetic and magnetic crystal materials \cite{TQCdata2019Vergniory, SymmIndi2019ZhangTT, SymmIndicatorData2019Tang, MTQCdata2020Xu,  GaojcNSR, wzjCDWTSM, PhysRevLett.124.076403QAHEuB6, PhysRevB.103.115145QSH235, GuozpQTI2022, PhysRevB.98.125143TaSe3},
a significant proportion of materials are topologically trivial, but host the type I unconventional feature of mismatch between the electronic Wannier centers and atomic positions, which are named as unconventional insulators or obstructed atomic insulators (OAIs) \cite{PhysRevB.103.205133electrides,niesimin_prr,GaojcUnconvMat,xu2021threedimensional}. 
Among them, various properties have been well studied already, such as electrides \cite{PhysRevB.103.205133electrides}, thermoelectronics \cite{GaojcUnconvMat}, catalysis \cite{guoweili}, electrochemistry \cite{interface1_usage}, solid-state hydrogen storage \cite{GaojcUnconvMat}, ferroelectronics and superconductivity \cite{ferroelecNature,supr-con1_usage, supr-con2_usage}.
Given an OAI with cleavage terminations cutting through an obstructed Wannier charge center, metallic surface states could possibly exist within the bulk gap, which is referred to as the filling anomaly \cite{multipoleScience, PhysRevB.96.245115multipole, PhysRevResearch.1.033074fractionalCorner, xu2021fillingenforced}.
Such metallic states on the surface/interface are key to many interesting physical and chemical properties of these unconventional materials \cite{GaojcUnconvMat, xu2021threedimensional}.
On the other hand, the concept of unconventionality can be extended to phonon spectra, when phonon modes are well separated in frequency (by a frequency gap) \cite{zhang2023unconventional}. 
For an unconventional phonon band structure, the surface phonon modes are expected in the frequency gap. So far, no material has been predicted to be unconventional yet in both electronic and phonon band structures.

In this work, we demonstrate that the trigonal selenium and tellurium are unconventional in both electronic and phonon spectra.
The essential band representation is $A@3b$ ($3b$ is an empty site). 
The real-space invariant (RSI) is computed $\delta_1@3b=-1$.
The $\pi$ Zak phase suggests that the 1D Se/Te chain is a chiral Su-Schrieffer-Heeger (SSH) chain.
The electronic gap can be modified by quantum-well structures, with electrons and holes being at two different interfaces, which yields a large shift current.
On the other hand, in their phonon bands, the band representations are assigned to three sets of separated phonon bands by the IR2PH \cite{ir2ph} and UnconvMat packages \cite{GaojcUnconvMat, unconvmat, GaojcIRVSP}. 
Accordingly, the obstructed surface phonon modes are obtained on the (0001) surface. Se and Te are the prototypes of unconventional materials in both electronic and phonon band structures.

\begin{table}[!b]
    \caption{
    The atomic positions and aBRs of SG 152 ($P3_{1}21$) Se and Te.
    }
    \label{tab:SeTe_aBR}
    \begin{ruledtabular}
    \begin{tabular}{c|c|c|c|c|c|c}
        Atom & WKS($q$) & Symm. & Conf. & Irreps($\rho$) & aBRs($\rho@q$) & occ. \\
        \hline
        Se & $3a$ & $2$ & $4p^{4}$ & $p_{x}$ :\: $A$ & $A@3a$ & \\
        (Te)& & & ($5p^{4}$) & $p_{y}$ :\: $B$ & $B@3a$ & yes \\
        & & & & $p_{z}$ :\: $B$ & $B@3a$ & \\
        \hline
        \multicolumn{5}{r}{} & {$A@3b$} & yes
    \end{tabular}
    \end{ruledtabular}
\end{table}

\paragraph*{Crystal structures and aBR analysis.}
\begin{figure}[!t]
    \centering
    \includegraphics[width=0.98\linewidth]{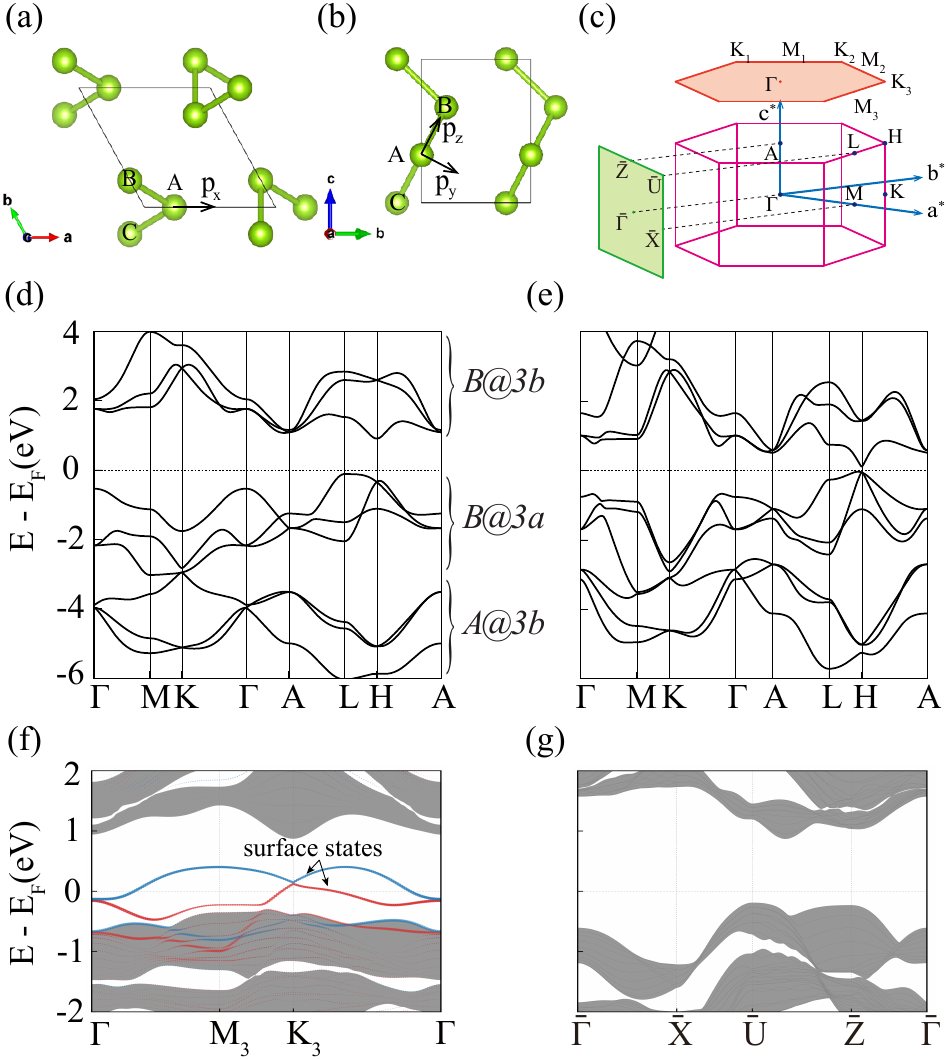}
    \caption{
    The crystal structure and Brillouin zone of trigonal Se and Te.
    (a) Top view and (b) side view of the crystal structure are presented.
    The black arrows denote the orientations of defined local orbitals according to $3a$ site symmetry.
    Each unit cell contains a chiral chain along $\hatc$ ($\hatz$).
    (c) The Brillouin zone and high-symmetry $k$ points of trigonal Se.
    (d),(e) The band structures of Se and Te.
    The obtained spectra of the (f) (0001)-slab and (g) ($0\bar{1}10$)-slab calculations of Se.
    The two surface bands are indicated by the red and blue circles, respectively, in panel (f).
    }
    \label{fig:dft}
\end{figure}

The Se and Te crystals crystallize in the structure of SG 152 (right-handed). 
The atoms are located the $3a$ Wyckoff site (WKS; \tab{tab:SeTe_aBR}), and covalently bonded into chiral 1D chains along $\hatz$. 
The neighboring chains are bonded by van der Waals (vdW) forces in the other two dimensions (\ie $xy$ plane) to form a triangular lattice [\fig{fig:dft}(a)]. 
The band structures of Se and Te are presented in Figs. \ref{fig:dft}(d) and \ref{fig:dft}(e), indicating an insulating behavior.
For more calculation details, please see Appendix \ref{sup:A}.
In the theory of TQC, one can identify the BR ($\rho@q$) for any isolated set of electronic bands, which tells the Wannier center ($q$) and local orbital character ($\rho$; labeled by $q$ site symmetry)~\cite{PhysRevB.103.205133electrides}. 
The atomic valence-electron BRs (aBRs) are tabulated in \tab{tab:SeTe_aBR}.
Based on the computed irreducible representations, the aBR decomposition for the occupied bands is solved online \cite{GaojcUnconvMat, unconvmat} to be $B@3a(p_{y}) + {A@3b}$, with the essential BR ${A@3b}$ responsible for charge mismatch.
The RSI for the empty $3b$ site is computed to be $\delta_{1}@3b \equiv -m(A) + m(B) = -1$, where $m(\rho)$ denotes the number of $\rho@3b$, indicating the unconventional nature of the OAI.
The specific orbital-resolved band structures and TB model are provided in Appendix \ref{app_sk-tb}.

\paragraph*{$\pi$ Zak phase and surface states.}
With the understanding of the essential BR at $3b$, we will expect the metallic state on the top surface while no such state on the side surface.
Thus, we have performed the (0001)- and ($0\bar{1}10$)-slab calculations.
In \fig{fig:dft}(f), we clearly see metallic surface states on the (0001) surface.
Moreover, we notice that the unconventional nature is closely related to the Zak phase~\cite{zak}. The Zak phase is defined by the integral of the Berry connection along a reciprocal lattice vector ($\vG$). In trigonal Se crystals, the Zak phase is defined by 
\begin{equation}
    \begin{aligned}
        \theta(\vk_{\parallel}) \eq -\ii \sum_{n=1}^{n_{\text{occ}}}
        \int_{0}^{\vG} \dd{k_{z}} 
        \mel{u_{n}(\vk_{\parallel}, k_{z})}{\partial_{k_{z}}}{u_{n}(\vk_{\parallel}, k_{z})}
    \end{aligned}
\end{equation}
where $\ket{u_{n}(\vk)}$ is the periodic part of the bulk Bloch wave function in the $n$th band,
with the gauge choice $\ket{u_{n}(\vk)} = e^{\ii\vG\cdot\br} \ket{u_{n}(\vk+\vG)}$ and $k_{\parallel}\equiv (k_{x},k_{y})$. 
Due to the presence of $\mathrm{C_{2x}}$ symmetry on ${\Gamma}$ - K/K - M and $\mathrm{TC_{2x}}$ symmetry on ${\Gamma}$ - M, the Zak phase is quantized to be zero or $\pi$(mod $2\pi$).
The Zak phase along ${\Gamma}$MK${\Gamma}$ is computed, being $\pi$.
Away from the high-symmetry lines, it would slightly vibrate around $\pi$.
The $\pi/0$ values of the Zak phase correspond to the presence/absence of in-gap boundary states, which are similar to those in the SSH model \cite{zakphase} . 
The 1D Se chains are chiral SSH chains.
As shown for the (0001) slab, polarization charges are equal to $e/2$ (mod $e$) at the ends of the 1D insulating chain. 
As these chiral SSH chains are vdW interacted, they result in metallic surface states, being consistent with \textit{ab-initio} calculations.

\begin{figure}[!t]
	\centering
	\includegraphics[width=0.98\linewidth]{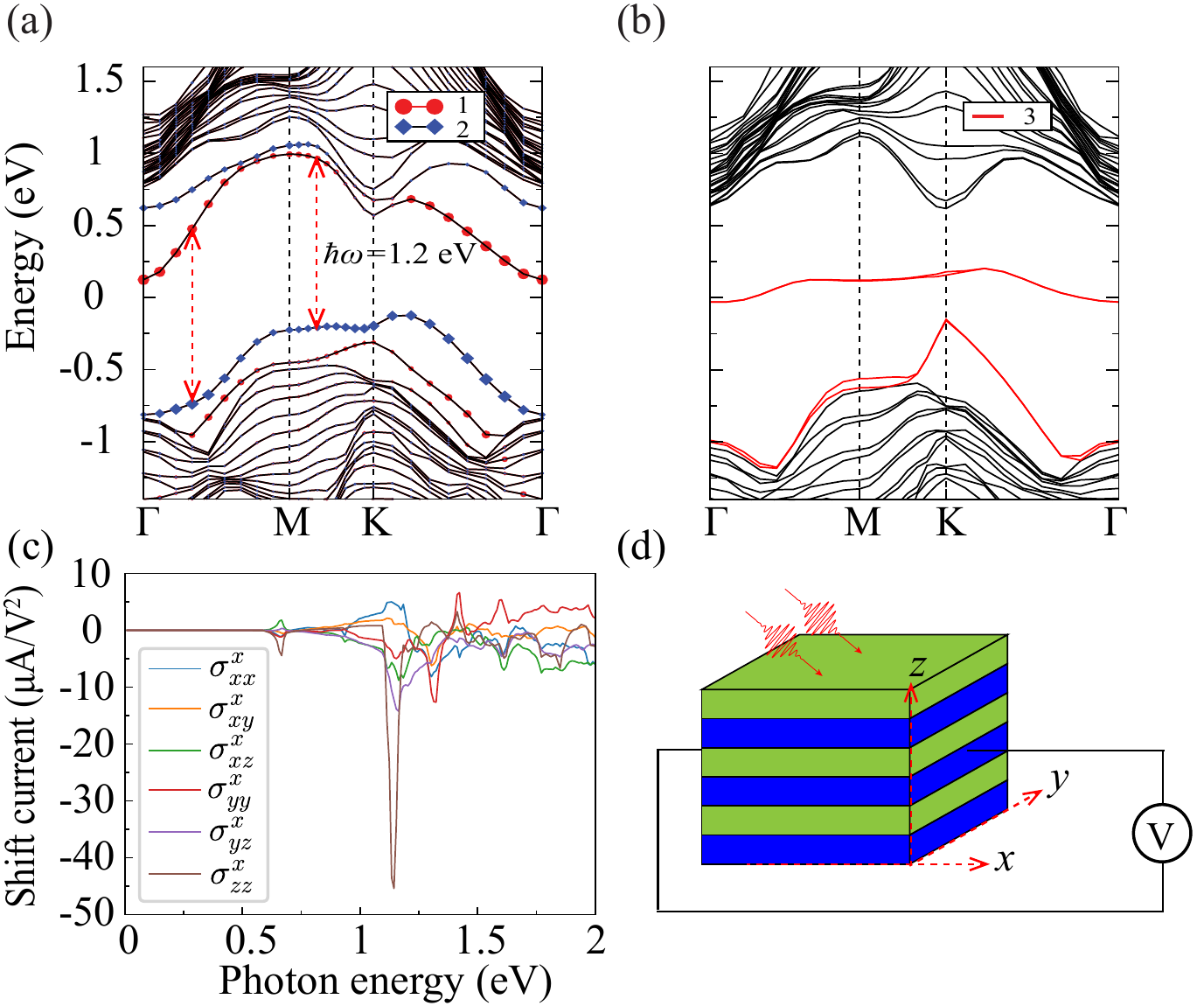}
	\caption{
		The band structures [(a),(b)] of the quantum wells of trigonal Se.  
		The in-gap interface states are shown clearly with in the bulk gap; one can easily find that the bands from two neighbor Se layers with direct band gaps $\sim 1.2\eV$ in a large area [red dashed arrows in panel (a)].
		The interface weights have been denoted by the size of the red circle and blue diamond dots for interfaces 1 and 2 in panel (a). The flat in-gap interface states (red solid line)
		are obtained on interface 3 in panel (b).
		(c) The response tenser elements corresponding to the shift currents along $\hatx$.
		The peak value for the $\sigma^{x}_{zz}(\omega)$ components can reach up to $\sim 45 \; \mu \text{A}/\text{V}^{2}$, with a photon energy of $\sim 1.2 \eV$.
		(d) Experiment setup for detecting the $\sigma^{x}_{zz}$.
	}
	\label{fig:interface}
\end{figure}

\paragraph*{Quantum wells, interfaces and shift current.}
Due to existence of the $\pi$ Zak phase, we can generate in-gap states on different (0001) interfaces, to modify the band gap and properties though the quantum-well structures.
Here, we twisted two different chiral structures by 60 degrees to form two interfaces 1 and 2 and 180 degrees to form interfaces 3. We have performed the calculations of two quantum wells (QWs). QW-I contains interfaces 1 and 2, while QW-II has identical interface 3.  
The top and side views of the structures are given in \fig{app_fig:interface} in Appendix \ref{app_shiftCurrent}. 
The band structures of the quantum wells are shown in Figs. \ref{fig:interface}(a) and \ref{fig:interface}(b).
In \fig{fig:interface}(b), the flat in-gap interface states are also obtained on interface 3.
In \fig{fig:interface}(a), the interface states are clearly seen in the bulk gap, and their surface weights have been denoted by the size of the dots for interfaces 1 and 2, respectively. 
We find that the indirect band gap becomes smaller in the QW-I structure. 
More surprisingly, electron and hole carriers accumulate at different interfaces, which can prohibit the recombination of electrons and holes.
 
Thus, we have calculated the shift current of this quantum well.  
In second-order optical response, a shift current $J^{c}(\omega)$ can be generated by an injected linear polarized light via 
\begin{equation}
J^{c}(\omega) = \sigma^{c}_{ab}(\omega) E_{a}(\omega) E_{b}(-\omega),
\end{equation}
where $E_{a}(\omega)$ is electrical field along $\hata$ and the strength of the response can be described by the magnitude of second-order photoconductivity $\sigma^{c}_{ab}(\omega)$ \cite{sc1,sc2,sc3}.
The results show a large shift current at $E(\hbar\omega) \sim 1.2 \eV$ in \fig{fig:interface}.
The response tenser elements corresponding to the shift current along $\hatx$, $\haty$, and $\hatz$ are given in \figs{fig:interface}(c), \ref{app_fig:shiftcurrent}(a) and \ref{app_fig:shiftcurrent}(b), respectively.
The peak value for the $\sigma^{x}_{zz}(\omega)$ and $\sigma^{y}_{zz}(\omega)$ components can reach up to $\sim 45$ and $\sim 80 \; \mu \text{A}/\text{V}^{2}$, respectively, with a photon energy of $\sim 1.2 \eV$.
So large photoconductivity is comparable to the record of reported values in semiconductors \cite{m1,m2,m3,m4,m5,m6,m7,m8,m9}, making the Se quantum well a promising photodetection device.
We further analyzed the origin of the strong response.
From the electronic band structure, one can easily find that the bands from two different interfaces (1 and 2) with direct band gaps $\sim 1.2\eV$ in a large area [see \fig{fig:interface}(a)].
The optical transition between the interfaces 1 and 2  contributes to the peak value of photoconductivity and can generate a strong in-plane shift current via the tenser elements $\sigma^{x}_{zz}(\omega)$ and $\sigma^{y}_{zz}(\omega)$.
The details of the second-order optical conductivity are given in Appendix \ref{sup:A}.

\begin{figure}[!t]
	\centering
	\includegraphics[width=0.98\linewidth]{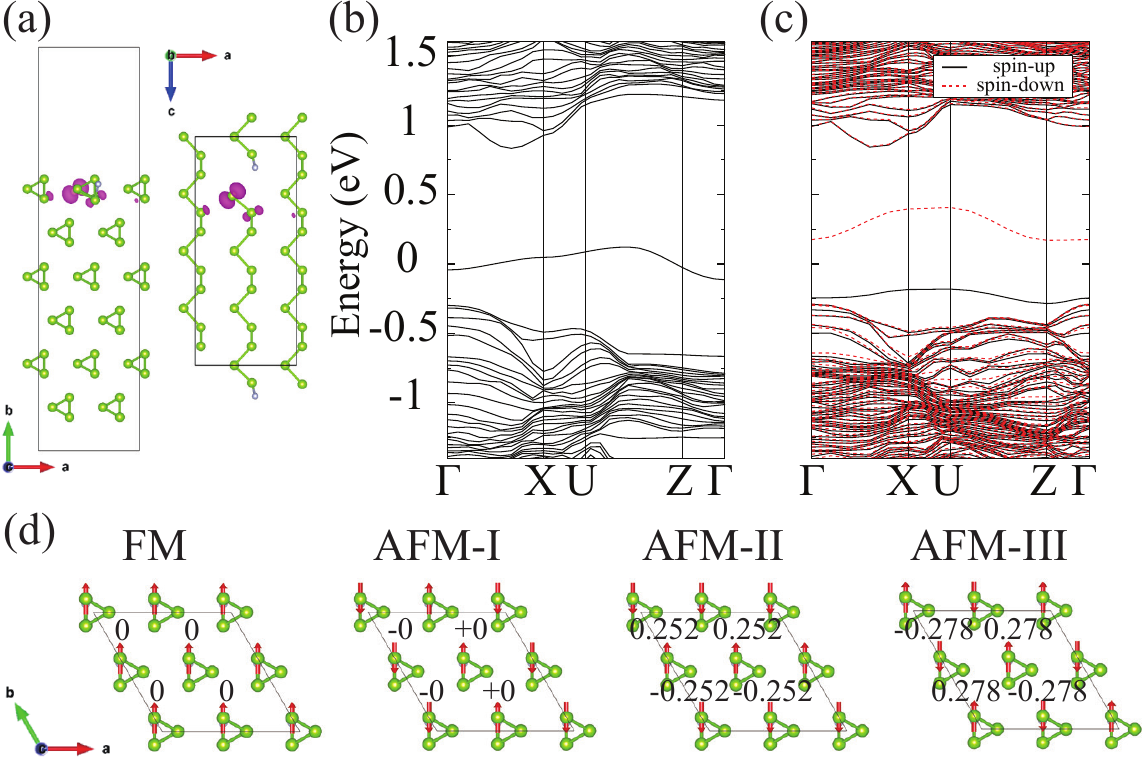}
	\caption{
		The structure of the decorated (0$\bar{1}10$) slab (a) and the computed band structures without (b) and with spin polarization (c). 
		Band decomposed charge density of the isolated in-gap band of (b) is plotted in (a) with an isosurface value of 0.01 $e$ \AA$^{-3}$.
		(d) The magnetic distribution of the $2\times2$ supercell of the (0001)-surface termination with different magnetic configurations. The arrows indicate the initial magnetic directions, while the values indicate the final converged magnetic moments.
	}
	\label{fig:mag}
\end{figure}

\paragraph*{Magnetism at the ends of the chiral SSH chain.}
To investigate the magnetic property at the ends of the chiral SSH chain, we induce a vacancy defect on the surface of the ($0\bar{1}10$) slab and passivate an end by the F atom. 
The relaxed crystal structure is obtained in \fig{fig:mag}(a). 
The non-spin-polarized and spin-polarized band structures are calculated. 
In the non-spin-polarized band structure of \fig{fig:mag}(b), an isolated state is localized at the end of the chain, being half filling and metallic.
The partial charge distribution of this end state is plotted in \fig{fig:mag}(a). 
After considering spin polarization, it becomes insulating with $1  \mu_B$ at the end with a lower total energy. The magnetism is induced by the unpaired $p$ electron at the end of a SSH chain. The polarized end state is confirmed by the calculations with two (or more) vacancies on a chain to reduce the interaction between the two ends of the chain in the slab structures.

Then we simulate the magnetic ordering on the ideal (0001) termination, which is an arrangement of the ends of the chiral SSH chains. 
We have performed the calculations for different magnetic configurations in a $2\times2$ supercell. 
The results are shown in \fig{fig:mag}(d).
We use the energy mapping method to estimate the nearest-neighbor exchange interaction. Here, we consider an Ising-type model,
\begin{equation}
    \begin{aligned}
        H \eq E_0+ \sum_{<i,j>}J_{i,j} S_{i} S_{j}.
    \end{aligned}
\end{equation}
Because there are three types of nearest-neighbor exchange interactions, the NN $J_{<i,j>}$ can be divided into $J_a$ along $\hata$, $J_{b}$ along $\hatb$, and $J_{ab}$ along $\hata+\hatb$. 
We consider one ferromagnetic (FM) and three antiferromagnetic (AFM) configurations, as indicated by the arrows in \fig{fig:mag}(d). 
We find that the FM and AFM-I states converge to paramagnetic states during the self-consistent calculations and the converged magnetic moments in AFM-II and AFM-III states are presented in \fig{fig:mag}(d).
The converged energies are given in \tab{table:mag_energy}.
According to the Ising-type model, the energies of the Se magnetic configurations can be written as,
\begin{equation}
    \begin{aligned}
        E_{\text{FM}} \eq E_0,\qquad  E_{\text{AFM-I}} = E_0, \\
        E_{\text{AFM-II}} \eq E_0 + (J_a-J_b-J_{ab})/4, \\
        E_{\text{AFM-III}} \eq E_0 + (-J_a-J_b+J_{ab})/4.
    \end{aligned}
\end{equation}
We can solve $J_b = 15.7 \meV$. 
Then, assuming $|J_a|=|J_{ab}|$, we get $J_a = -J_{ab} = 1.2 \meV$. 
As expected, $J_a/J_{ab}$ is much less then $J_b$ and these $J$ values are consistent with the most stable configuration AFM-III.

\begin{table}[!tbh]
	\caption{
	The energies of different magnetic configurations (reference to AFM-III).
	}
	\label{table:mag_energy}
	\begin{ruledtabular}
	\begin{tabular}{ccccc}
		 & FM & AFM-I & AFM-II & AFM-III  \\\hline 
		Energy (meV) & 72.38 & 72.38 & 19.08 & 0
	\end{tabular}
	\end{ruledtabular}
\end{table}

\begin{figure}[!t]
    \centering
    \includegraphics[width=0.98\linewidth]{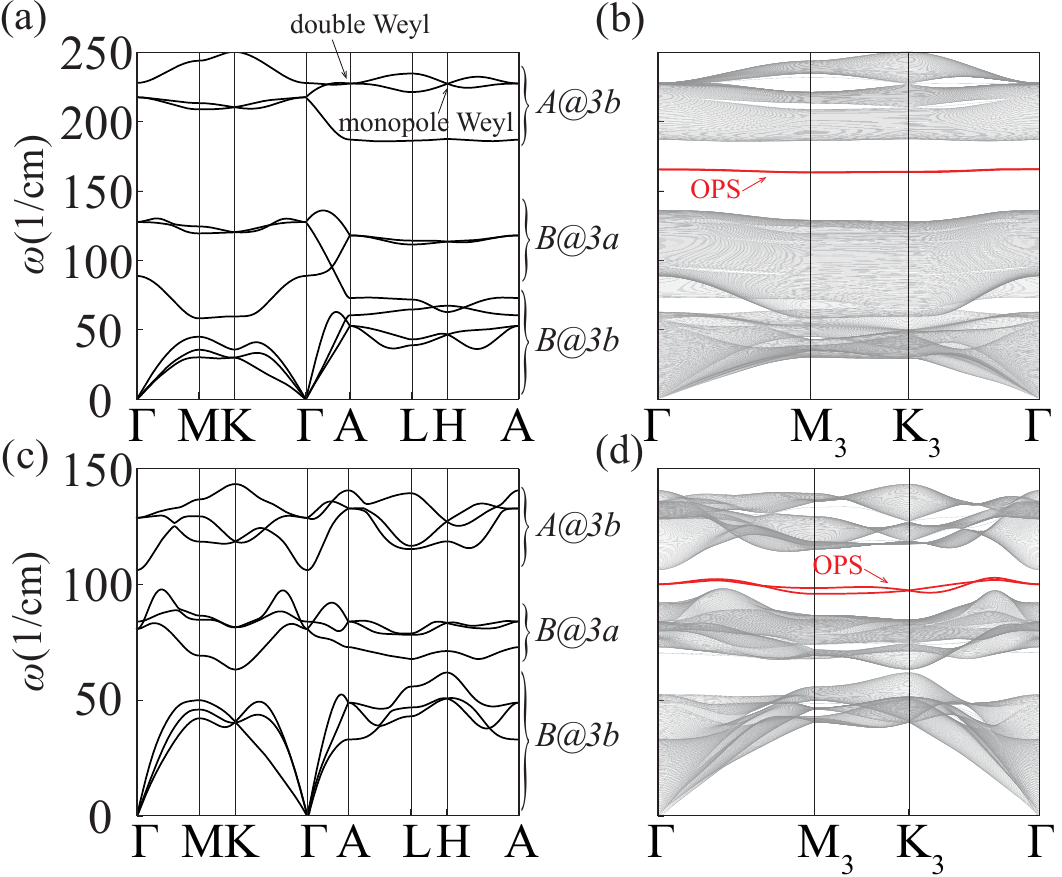}
    \caption{
    The calculated bulk and (0001)-surface phonon spectra of Se [(a,b)] and Te [(c,d)]. The obstructed phonon states (OPS) are clearly shown in the frequency gap of the bulk projections.}
    \label{fig:phonon_band}
\end{figure}

\paragraph*{Unconventional phonon spectra.}
The phonon modes are originated from the vibration of the atoms, which would induce the atomic vibration band representations as the p-orbitals in the phonon spectrum.
Similarly, the phonon spectra with a set of separated phonon bands being not a sum of aBRs are defined as the unconventional phonon spectra \cite{ zhang2023unconventional}. 
As a result, the obstructed phonon modes emerge on terminations of finite-size samples.
The phonon spectra are computed and shown in \fig{fig:phonon_band} for Se and Te respectively. We use IR2PH to deal with the phonon modes from the DFPT calculations and obtain the irreps \cite{ir2ph}. 
Through \texttt{tqc.data}, the BR decomposition for each set of phonon bands is solved online \cite{unconvmat}, as shown in Figs. \ref{fig:phonon_band}(a) and \ref{fig:phonon_band}(c). 
As expected, the obstructed phonon states are obtained in Figs. \ref{fig:phonon_band}(b) and \ref{fig:phonon_band}(d).
Based on the full list of symmetry-protected Weyl phonons in Ref. \cite{weyl_phonon}, the depicted A and H points are double and monopole Weyl points, respectively. The arc states of Weyl points are shown in \fig{app_fig:phSSs} in Appendix \ref{app_phononWeyl}.

\paragraph*{Discussion.}
In summary, we present that the trigonal Se and Te are unconventional elements in both electronic and phonon spectra.
The essential BR of the occupied electronic bands is $B@3b$ and it is characterized by a nonzero RSI $\delta@3b=-1$. 
Since $3b$ site is the center of bonds in the 1D chiral Se/Te chain in the $z$ direction ($\pi$ Zak phase), the metallic in-gap states are obtained on the (0001) surfaces and interfaces. 
Noting that the topological property of the Se/Te chain was investigated very recently \cite{chalcogen}.
As semiconductors, the band gap can be modified by the quantum-well structure with electrons and holes being at two different interfaces, which benefits to the application of the solar cells. 
A super-high shift current is obtained in the quantum well.
Additionally, based on the BR decomposition of the phonon bands, we conclude that the
phonon bands are unconventional and the obstructed surface phonon modes are obtained on the (0001) surface/interface.
As the prototypes of a novel kind of materials with both electronic and phonon unconventional band structures, our study can create future interest in the study of surface charge and phonons in these  kind of materials.

\paragraph*{Acknowledgments.}
We thank Dr. Qing-Bo Liu for finding the Weyl points and calculating surface arc states of the Se phonon spectrum. This work was supported by the National Key R\&D Program of China (Grant No. 2022YFA1403800), National Natural Science Foundation of China (Grants No. 11974395, No. 12188101 and No. 52188101), the Strategic Priority Research Program of Chinese Academy of Sciences (Grant No. XDB33000000), the China Postdoctoral Science Foundation funded project (Grant No. 2021M703461), and the Center for Materials Genome.

%


\beginsupplement{}

\section*{Appendix}

\subsection{\label{sup:A}CRYSTAL STRUCTURE AND CALCULATION METHODS}
\begin{figure}[htbp]
	\centering
	\includegraphics[width=0.9\linewidth]{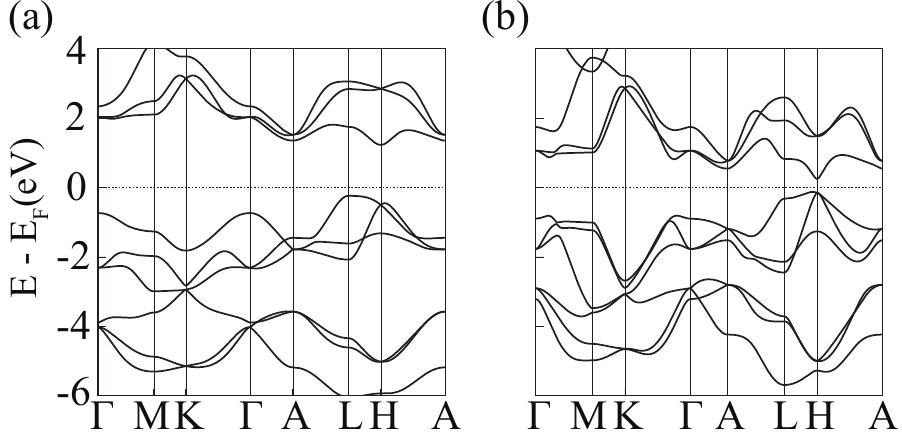}
	\caption{
		The MBJ band structures of Se and Te.
	}
	\label{app_fig:mbj}
\end{figure}

The elementary substances Se and Te (SG 152, $P3_121$) host a hexagonal lattice (parameters $a, c$ along $\hatx, \hatz$, respectively),
whose atoms located at Wyckoff positions $3a$, \ie $A = (x, 0, 1/3)$, $B = (0, x, 2/3)$, and $C = (-x, -x, 0)$, forming a right-handed trigonal helix along $\hatz$.
The parameters are $(a, c, x) = (4.4593\AAA, 5.9282\AAA, 0.26197)$ for Te and $(a, c, x) = (4.3662\AAA, 4.9536\AAA, 0.2254)$ for Se.
The modified Becke Johnson (MBJ) band structures of Se and Te are presented in Figs. \ref{app_fig:mbj}(a) and \ref{app_fig:mbj}(b) \cite{PhysRevLett.102mbj,J.Chem.Phys.124mbj} .

We performed the first-principles calculations based on the density functional theory (DFT) implemented in the Quantum ESPRESSO package \cite{Giannozzi_2009, Giannozzi_2017}, using the projector augmented-wave (PAW) method \cite{PhysRevB.50.17953PAW1, PhysRevB.59.1758PAW2}. 
The electrons are described using the generalized gradient approximation (GGA) with exchange-correlation functional of Perdew, Burke and Ernzerhof (PBE) \cite{GGA-PBE1996} for the exchange-correlation functional. 
The plane-wave cutoff energy was set to 70 Ry. The thickness of the vacuum layer along $\hatz$ axis was set to $>15$ \AA. The lattice dynamical properties are described using the GGA with exchange-correlation functional of PBE for the exchange-correlation functional. The plane-wave cutoff energy was set to 50 Ry. The phonon calculations are performed in the framework of density functional perturbation theory, as implemented in the Quantum ESPRESSO package.

The irreducible representations of electronic states and phonon modes were computed by our homemade programs IR2PW and IR2PH, respectively. They are released on Github \cite{ir2ph}.

	\begin{figure}[h]
		\centering
		\includegraphics[width=0.9\linewidth]{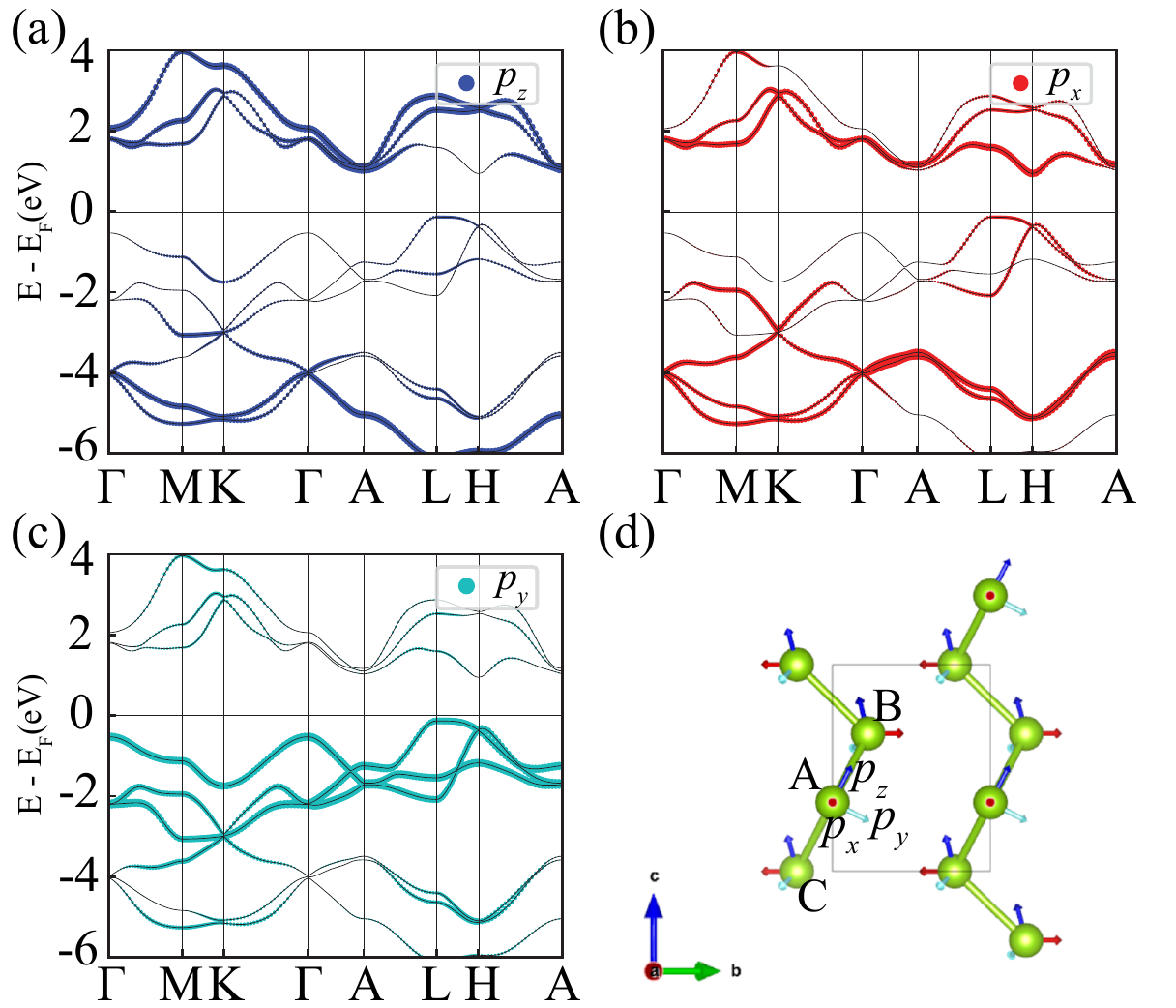}
		\caption{
			Orbital-resolved Wannier-band structures for Se; the size of the circles denotes the weight of defined local (a) $p_{z}$, (b) $p_{x}$, and (c) $p_{y}$ orbitals, respectively.
			(d) Illustration for the orientations of defined local orbitals (red arrows: $p_{x}$, cyan arrows: $p_{y}$, and blue arrows: $p_{z}$).
		}
		\label{app_fig:fat}
	\end{figure}

\begin{figure}[htb!p]
	\centering
	\includegraphics[width=0.9\linewidth]{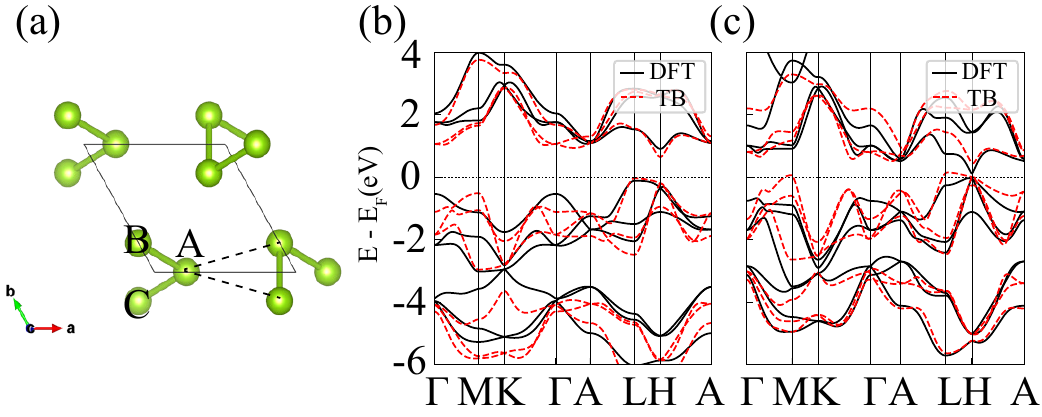}
	\caption{
		(a) The illustration of NN bonds (green solid lines) and NNN bonds (black dashed lines).
		The band structures of TB models (red dashed lines) v.s. \textit{ab-initio} calculations (black solid lines) of Se (b) and Te (c).
	}
	\label{app_fig:sk}
\end{figure}

The same band topology, we mainly focus on the discussion of Se in the main text hereafter.

It is physically different from that of the p-n junction mechanism. In bulk optical response, the linear response can be understood as
$J^{a}(\omega) = \sigma_{ab}(\omega) E_{b}(\omega)$, where $\sigma_{ab}(\omega)$ is the first-order optical conductivity. 
A current $J^{a}(\omega)$ along the a direction can be generated by the applied optical-electrical filed $E_{b}(\omega)$ along the b direction. 
One can see that $J^{a}(\omega)$ and $E_{b}(\omega)$ are with the same frequency; therefore, 
the bulk photovoltaic effect is forbidden in the first-order response and dominated by the second-order response. The bulk photovoltaic effect can be considered as 
a joint effect of the frequency-dependent density of states and the wave function.

We calculate the second-order optical conductivity by
\begin{equation}
	\begin{aligned}
	J^{c}(0) = &\sigma^{c}_{ab}(0,\omega,-\omega) E_{a}(\omega) E_{b}(-\omega)\\
	&\times Re[E_{a}(\omega)E_{b}(-\omega)] \\
	\sigma ^{c}_{ab}(0,\omega,-\omega) = &\frac{ie^{3}\pi}{\hbar V}\sum_{k}\sum_{m,n}f_{nm}(r_{mn}^{a}r_{nm}^{b;a}+r_{mn}^{b}r_{nm}^{b;c})\\
	&\times \delta(\hbar\omega-E_{mn})
    \end{aligned}
\end{equation}

where $n$ and $m$ are band indices, $f_{nm}=f_{n}-f_{m}$ is the Fermi-Dirac distribution difference, $r_{mn}^{a}=iA_{mn}^{a}(1-\delta_{nm})$ 
is the interband Berry connection, $r_{nm}^{a;b}=\frac{\partial{r_{nm}^{a}}}{\partial{k_{b}}}$
the term relates to the mediate virtual states, and $E_{nm}=E_{n}-E_{m}$ is the energy difference. We deal with the delta
function $\delta(\hbar\omega-E_{mn})$ by a Gauss distribution with a smearing factor of 0.01eV.

\subsection{\label{app_sk-tb}ORBITAL-RESOLVED BAND STRUCTURES AND TB MODEL}

The orientations of the defined local orbitals at each Se atoms according to $3b$ site symmetry are presented in \fig{app_fig:fat}(d).
The orbital-resolved band with respect to this orbital setup are shown in Figs. \ref{app_fig:fat}(a)--(c), from which we can find that the defined $p_{y}$ orbitals are fully occupied forming the aBR $B@3a$.

Though the Slater-Koster method \cite{slater-koster}, the band structures can be well reproduced by a $p-$orbital tight-binding (TB) model with two kinds of bonds: 
the nearest-neighbor (NN) bonds and the next-nearest-neighbor (NNN) bonds in \fig{app_fig:sk}.
Their $pp\sigma$ and $pp\pi$ parameters are given in  \tab{app_tab:sk}.
The band structures of the TB model are given in in \fig{app_fig:sk}.

\begin{table}[htb!p]
    \caption{
    The Slater-Koster parameters of the $p-$orbital TB model of Se and Te. 
    $V_{pp\sigma/pp\pi}$ ($V'_{pp\sigma/pp\pi}$) denote the parameters of NN (NNN) bonds.
    }
    \label{app_tab:sk}
    \begin{ruledtabular}
    \begin{tabular}{ccccc}
        Atom & $V_{pp\sigma}$ & $V_{pp\pi}$ & $V'_{pp\sigma}$ & $V'_{pp\pi}$ \\\hline
        Se & $3.2$ & $-0.5$ & $1.2$ & $-0.3$ \\\
        Te & $2.5$ & $-0.2$ & $1.3$ & $-0.285$
    \end{tabular}
    \end{ruledtabular}
\end{table}

\subsection{\label{app_shiftCurrent}  LARGE SHIFT CURRENT AND QUANTUM WELLS}
\begin{figure}[htbp]
	\centering
	\includegraphics[width=0.9\linewidth]{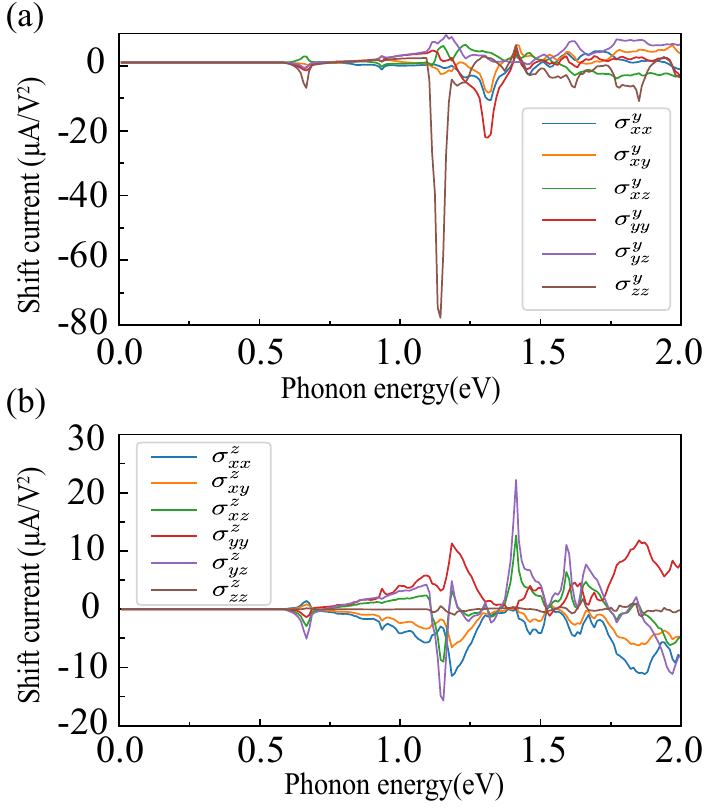}
	\caption{\label{app_fig:shiftcurrent}
		(a) and (b) The response tenser elements corresponding to the shift currents along $\haty$, and $\hatz$.
		The peak value for the $\sigma^{y}_{zz}(\omega)$ components can reach up to $\sim 80 \; \mu \text{A}/\text{V}^{2}$, with a photon energy of $\sim 1.2 \eV$.
	}
\end{figure}

\begin{figure}[htbp]
	\centering
	\includegraphics[width=0.9\linewidth]{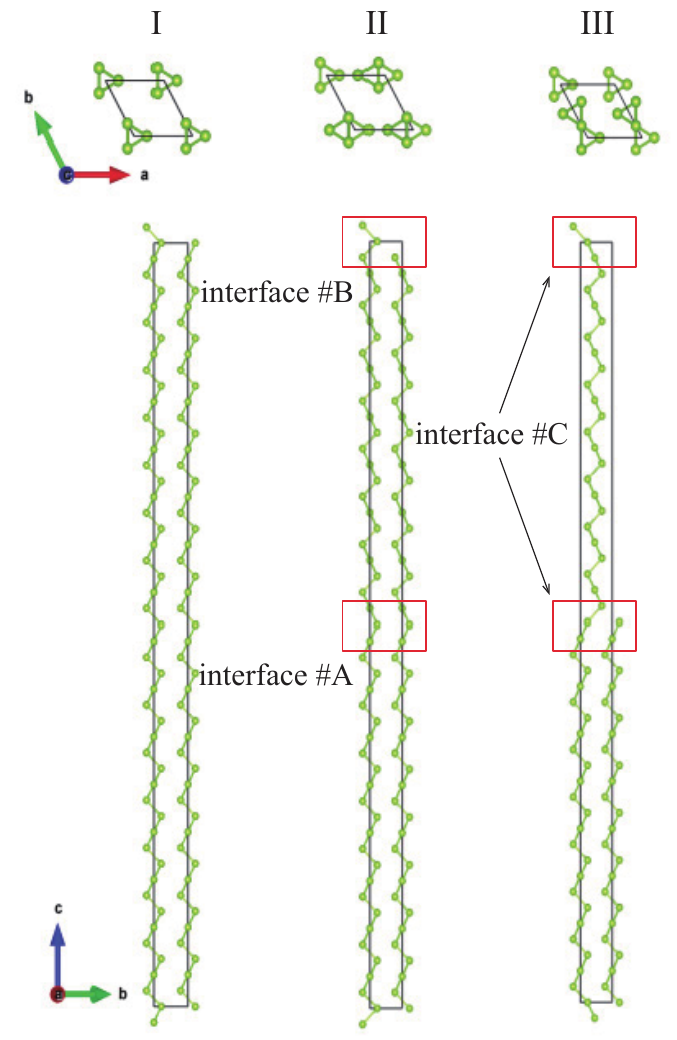}
	\caption{\label{app_fig:interface}
		The crystal structures of the quantum wells of trigonal Se. 
		QW-I has two interfaces (1 and 2) and QW-II has two identical interfaces (3). 
	}
\end{figure}

These topics are covered in Figs. \ref{app_fig:shiftcurrent} and \ref{app_fig:interface}.

\clearpage

\subsection{\label{app_phononWeyl}WEYL POINTS AND SURFACE STATES IN PHONON SPECTRUM}
\begin{figure}[!h]
	\includegraphics[width=0.9\linewidth]{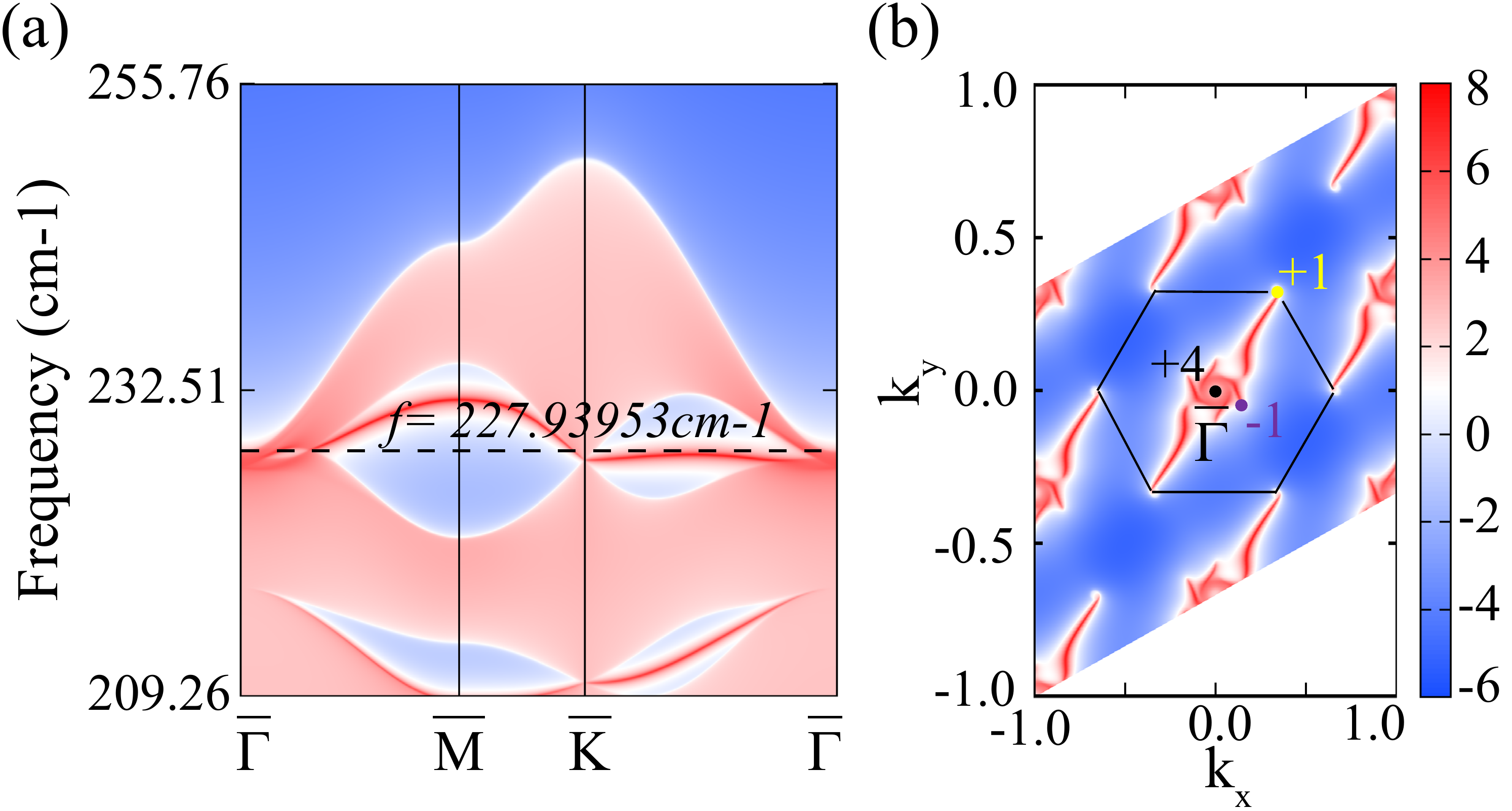}
	\caption{\label{app_fig:phSSs}
		(a) The (0001) surface phonon spectrum of a semi-infinite Se.
		(b) The surface arc states terminated by the projection [dash line in (a)] of Weyl nodes with unequal topological charge.
		The yellow and purple points denote the Weyl nodes with chiral charge $+1$ and $-1$, respectively.
		Two charge $+1$ and one charge $+2$ Weyl nodes at $\bar{\Gamma}$ degenerate to $+4$ Weyl nodes, resulting in quadruple helical states and surface arcs.
	}
\end{figure}

The Weyl points between 8th and 9th phonon bands in the phonon spectrum of Se are listed in \tab{app_tab:wp}.
The corresponding surface/arc states are obtained in \fig{app_fig:phSSs} for (0001) surface.

\begin{table}[!h]
	\caption{\label{app_tab:wp} The location, frequency, chiral charge, and degeneracy of the Weyl points between 8th and 9th phonon bands.
	}
	\begin{ruledtabular}
		\begin{tabular}{c|c|c|c|c}
			Weyl points & Loc. (fractional) & Freq. ($\text{cm}^{-1}$) & Charge & Deg. \\
			\hline
			WP1 & $(0.0,0.0,0.5)$ & $227.34896$ & $+2$ & $1$ \\
			
			WP2 & $(0.0,0.0,0.24)$ & $226.65376$ & $+1$ & $2$ \\
			
			WP3 & $(0.33,0.33,0.5)$ & $227.16528$ & $+1$ & $2$ \\
			
			WP4 & $(0.0, -0.15, -0.35)$ & $227.93953$ & $-1$ & $6$ \\
		\end{tabular}
	\end{ruledtabular}
\end{table}


\begin{thebibliography}{59}%
\makeatletter
\providecommand \@ifxundefined [1]{%
 \@ifx{#1\undefined}
}%
\providecommand \@ifnum [1]{%
 \ifnum #1\expandafter \@firstoftwo
 \else \expandafter \@secondoftwo
 \fi
}%
\providecommand \@ifx [1]{%
 \ifx #1\expandafter \@firstoftwo
 \else \expandafter \@secondoftwo
 \fi
}%
\providecommand \natexlab [1]{#1}%
\providecommand \enquote  [1]{``#1''}%
\providecommand \bibnamefont  [1]{#1}%
\providecommand \bibfnamefont [1]{#1}%
\providecommand \citenamefont [1]{#1}%
\providecommand \href@noop [0]{\@secondoftwo}%
\providecommand \href [0]{\begingroup \@sanitize@url \@href}%
\providecommand \@href[1]{\@@startlink{#1}\@@href}%
\providecommand \@@href[1]{\endgroup#1\@@endlink}%
\providecommand \@sanitize@url [0]{\catcode `\\12\catcode `\$12\catcode
  `\&12\catcode `\#12\catcode `\^12\catcode `\_12\catcode `\%12\relax}%
\providecommand \@@startlink[1]{}%
\providecommand \@@endlink[0]{}%
\providecommand \url  [0]{\begingroup\@sanitize@url \@url }%
\providecommand \@url [1]{\endgroup\@href {#1}{\urlprefix }}%
\providecommand \urlprefix  [0]{URL }%
\providecommand \Eprint [0]{\href }%
\providecommand \doibase [0]{http://dx.doi.org/}%
\providecommand \selectlanguage [0]{\@gobble}%
\providecommand \bibinfo  [0]{\@secondoftwo}%
\providecommand \bibfield  [0]{\@secondoftwo}%
\providecommand \translation [1]{[#1]}%
\providecommand \BibitemOpen [0]{}%
\providecommand \bibitemStop [0]{}%
\providecommand \bibitemNoStop [0]{.\EOS\space}%
\providecommand \EOS [0]{\spacefactor3000\relax}%
\providecommand \BibitemShut  [1]{\csname bibitem#1\endcsname}%
\let\auto@bib@innerbib\@empty
\bibitem [{\citenamefont {Bradlyn}\ \emph {et~al.}(2017)\citenamefont {Bradlyn}
  \emph {et~al.}}]{TQCBradlyn2017}%
  \BibitemOpen
  \bibfield  {author} {\bibinfo {author} {\bibfnamefont {Barry}\ \bibnamefont
  {Bradlyn}} \emph {et~al.},\ }\bibfield  {title} {\enquote {\bibinfo {title}
  {Topological quantum chemistry},}\ }\href {\doibase 10.1038/nature23268}
  {\bibfield  {journal} {\bibinfo  {journal} {Nature}\ }\textbf {\bibinfo
  {volume} {547}},\ \bibinfo {pages} {298--305} (\bibinfo {year}
  {2017})}\BibitemShut {NoStop}%
\bibitem [{\citenamefont {Elcoro}\ \emph {et~al.}(2021)\citenamefont {Elcoro},
  \citenamefont {Wieder}, \citenamefont {Song}, \citenamefont {Xu},
  \citenamefont {Bradlyn},\ and\ \citenamefont {Bernevig}}]{MTQC2021Luis}%
  \BibitemOpen
  \bibfield  {author} {\bibinfo {author} {\bibfnamefont {Luis}\ \bibnamefont
  {Elcoro}}, \bibinfo {author} {\bibfnamefont {Benjamin~J.}\ \bibnamefont
  {Wieder}}, \bibinfo {author} {\bibfnamefont {Zhida}\ \bibnamefont {Song}},
  \bibinfo {author} {\bibfnamefont {Yuanfeng}\ \bibnamefont {Xu}}, \bibinfo
  {author} {\bibfnamefont {Barry}\ \bibnamefont {Bradlyn}}, \ and\ \bibinfo
  {author} {\bibfnamefont {B.~Andrei}\ \bibnamefont {Bernevig}},\ }\bibfield
  {title} {\enquote {\bibinfo {title} {Magnetic topological quantum
  chemistry},}\ }\href {\doibase 10.1038/s41467-021-26241-8} {\bibfield
  {journal} {\bibinfo  {journal} {Nat. Commun.}\ }\textbf {\bibinfo {volume}
  {12}},\ \bibinfo {pages} {5965} (\bibinfo {year} {2021})}\BibitemShut
  {NoStop}%
\bibitem [{\citenamefont {Gao}\ \emph {et~al.}(2022{\natexlab{a}})\citenamefont
  {Gao}, \citenamefont {Guo}, \citenamefont {Weng},\ and\ \citenamefont
  {Wang}}]{GaojcMagUnconvMat}%
  \BibitemOpen
  \bibfield  {author} {\bibinfo {author} {\bibfnamefont {Jiacheng}\
  \bibnamefont {Gao}}, \bibinfo {author} {\bibfnamefont {Zhaopeng}\
  \bibnamefont {Guo}}, \bibinfo {author} {\bibfnamefont {Hongming}\
  \bibnamefont {Weng}}, \ and\ \bibinfo {author} {\bibfnamefont {Zhijun}\
  \bibnamefont {Wang}},\ }\bibfield  {title} {\enquote {\bibinfo {title}
  {Magnetic band representations, fu-kane-like symmetry indicators, and
  magnetic topological materials},}\ }\href {\doibase
  10.1103/PhysRevB.106.035150} {\bibfield  {journal} {\bibinfo  {journal}
  {Phys. Rev. B}\ }\textbf {\bibinfo {volume} {106}},\ \bibinfo {pages}
  {035150} (\bibinfo {year} {2022}{\natexlab{a}})}\BibitemShut {NoStop}%
\bibitem [{\citenamefont {Po}\ \emph {et~al.}(2017)\citenamefont {Po},
  \citenamefont {Vishwanath},\ and\ \citenamefont
  {Watanabe}}]{SymmIndicator2017Po}%
  \BibitemOpen
  \bibfield  {author} {\bibinfo {author} {\bibfnamefont {Hoi~Chun}\
  \bibnamefont {Po}}, \bibinfo {author} {\bibfnamefont {Ashvin}\ \bibnamefont
  {Vishwanath}}, \ and\ \bibinfo {author} {\bibfnamefont {Haruki}\ \bibnamefont
  {Watanabe}},\ }\bibfield  {title} {\enquote {\bibinfo {title} {Symmetry-based
  indicators of band topology in the 230 space groups},}\ }\href {\doibase
  10.1038/s41467-017-00133-2} {\bibfield  {journal} {\bibinfo  {journal} {Nat.
  Commun.}\ }\textbf {\bibinfo {volume} {8}},\ \bibinfo {pages} {50} (\bibinfo
  {year} {2017})}\BibitemShut {NoStop}%
\bibitem [{\citenamefont {Watanabe}\ \emph {et~al.}(2018)\citenamefont
  {Watanabe}, \citenamefont {Po},\ and\ \citenamefont
  {Vishwanath}}]{MSymmIndicator2018Po}%
  \BibitemOpen
  \bibfield  {author} {\bibinfo {author} {\bibfnamefont {Haruki}\ \bibnamefont
  {Watanabe}}, \bibinfo {author} {\bibfnamefont {Hoi~Chun}\ \bibnamefont {Po}},
  \ and\ \bibinfo {author} {\bibfnamefont {Ashvin}\ \bibnamefont
  {Vishwanath}},\ }\bibfield  {title} {\enquote {\bibinfo {title} {Structure
  and topology of band structures in the 1651 magnetic space groups},}\ }\href
  {\doibase 10.1126/sciadv.aat8685} {\bibfield  {journal} {\bibinfo  {journal}
  {Sci. Adv.}\ }\textbf {\bibinfo {volume} {4}},\ \bibinfo {pages} {eaat8685}
  (\bibinfo {year} {2018})}\BibitemShut {NoStop}%
\bibitem [{\citenamefont {Song}\ \emph {et~al.}(2018)\citenamefont {Song},
  \citenamefont {Zhang}, \citenamefont {Fang},\ and\ \citenamefont
  {Fang}}]{SymmIndi2018Song}%
  \BibitemOpen
  \bibfield  {author} {\bibinfo {author} {\bibfnamefont {Zhida}\ \bibnamefont
  {Song}}, \bibinfo {author} {\bibfnamefont {Tiantian}\ \bibnamefont {Zhang}},
  \bibinfo {author} {\bibfnamefont {Zhong}\ \bibnamefont {Fang}}, \ and\
  \bibinfo {author} {\bibfnamefont {Chen}\ \bibnamefont {Fang}},\ }\bibfield
  {title} {\enquote {\bibinfo {title} {Quantitative mappings between symmetry
  and topology in solids},}\ }\href {\doibase 10.1038/s41467-018-06010-w}
  {\bibfield  {journal} {\bibinfo  {journal} {Nat. Commun.}\ }\textbf {\bibinfo
  {volume} {9}},\ \bibinfo {pages} {3530} (\bibinfo {year} {2018})}\BibitemShut
  {NoStop}%
\bibitem [{\citenamefont {Kruthoff}\ \emph {et~al.}(2017)\citenamefont
  {Kruthoff}, \citenamefont {de~Boer}, \citenamefont {van Wezel}, \citenamefont
  {Kane},\ and\ \citenamefont {Slager}}]{PhysRevX.7.041069SymmIndicator}%
  \BibitemOpen
  \bibfield  {author} {\bibinfo {author} {\bibfnamefont {Jorrit}\ \bibnamefont
  {Kruthoff}}, \bibinfo {author} {\bibfnamefont {Jan}\ \bibnamefont {de~Boer}},
  \bibinfo {author} {\bibfnamefont {Jasper}\ \bibnamefont {van Wezel}},
  \bibinfo {author} {\bibfnamefont {Charles~L.}\ \bibnamefont {Kane}}, \ and\
  \bibinfo {author} {\bibfnamefont {Robert-Jan}\ \bibnamefont {Slager}},\
  }\bibfield  {title} {\enquote {\bibinfo {title} {Topological classification
  of crystalline insulators through band structure combinatorics},}\ }\href
  {\doibase 10.1103/PhysRevX.7.041069} {\bibfield  {journal} {\bibinfo
  {journal} {Phys. Rev. X}\ }\textbf {\bibinfo {volume} {7}},\ \bibinfo {pages}
  {041069} (\bibinfo {year} {2017})}\BibitemShut {NoStop}%
\bibitem [{\citenamefont {Peng}\ \emph {et~al.}(2022)\citenamefont {Peng},
  \citenamefont {Jiang}, \citenamefont {Fang}, \citenamefont {Weng},\ and\
  \citenamefont {Fang}}]{peng2021topological}%
  \BibitemOpen
  \bibfield  {author} {\bibinfo {author} {\bibfnamefont {Bingrui}\ \bibnamefont
  {Peng}}, \bibinfo {author} {\bibfnamefont {Yi}~\bibnamefont {Jiang}},
  \bibinfo {author} {\bibfnamefont {Zhong}\ \bibnamefont {Fang}}, \bibinfo
  {author} {\bibfnamefont {Hongming}\ \bibnamefont {Weng}}, \ and\ \bibinfo
  {author} {\bibfnamefont {Chen}\ \bibnamefont {Fang}},\ }\bibfield  {title}
  {\enquote {\bibinfo {title} {Topological classification and diagnosis in
  magnetically ordered electronic materials},}\ }\href {\doibase
  10.1103/PhysRevB.105.235138} {\bibfield  {journal} {\bibinfo  {journal}
  {Phys. Rev. B}\ }\textbf {\bibinfo {volume} {105}},\ \bibinfo {pages}
  {235138} (\bibinfo {year} {2022})}\BibitemShut {NoStop}%
\bibitem [{\citenamefont {Vergniory}\ \emph {et~al.}(2019)\citenamefont
  {Vergniory}, \citenamefont {Elcoro}, \citenamefont {Felser}, \citenamefont
  {Regnault}, \citenamefont {Bernevig},\ and\ \citenamefont
  {Wang}}]{TQCdata2019Vergniory}%
  \BibitemOpen
  \bibfield  {author} {\bibinfo {author} {\bibfnamefont {M.~G.}\ \bibnamefont
  {Vergniory}}, \bibinfo {author} {\bibfnamefont {L.}~\bibnamefont {Elcoro}},
  \bibinfo {author} {\bibfnamefont {Claudia}\ \bibnamefont {Felser}}, \bibinfo
  {author} {\bibfnamefont {Nicolas}\ \bibnamefont {Regnault}}, \bibinfo
  {author} {\bibfnamefont {B.~Andrei}\ \bibnamefont {Bernevig}}, \ and\
  \bibinfo {author} {\bibfnamefont {Zhijun}\ \bibnamefont {Wang}},\ }\bibfield
  {title} {\enquote {\bibinfo {title} {A complete catalogue of high-quality
  topological materials},}\ }\href {\doibase 10.1038/s41586-019-0954-4}
  {\bibfield  {journal} {\bibinfo  {journal} {Nature}\ }\textbf {\bibinfo
  {volume} {566}},\ \bibinfo {pages} {480--485} (\bibinfo {year}
  {2019})}\BibitemShut {NoStop}%
\bibitem [{\citenamefont {Zhang}\ \emph {et~al.}(2019)\citenamefont {Zhang}
  \emph {et~al.}}]{SymmIndi2019ZhangTT}%
  \BibitemOpen
  \bibfield  {author} {\bibinfo {author} {\bibfnamefont {Tiantian}\
  \bibnamefont {Zhang}} \emph {et~al.},\ }\bibfield  {title} {\enquote
  {\bibinfo {title} {Catalogue of topological electronic materials},}\ }\href
  {\doibase 10.1038/s41586-019-0944-6} {\bibfield  {journal} {\bibinfo
  {journal} {Nature}\ }\textbf {\bibinfo {volume} {566}},\ \bibinfo {pages}
  {475--479} (\bibinfo {year} {2019})}\BibitemShut {NoStop}%
\bibitem [{\citenamefont {Tang}\ \emph {et~al.}(2019)\citenamefont {Tang},
  \citenamefont {Po}, \citenamefont {Vishwanath},\ and\ \citenamefont
  {Wan}}]{SymmIndicatorData2019Tang}%
  \BibitemOpen
  \bibfield  {author} {\bibinfo {author} {\bibfnamefont {Feng}\ \bibnamefont
  {Tang}}, \bibinfo {author} {\bibfnamefont {Hoi~Chun}\ \bibnamefont {Po}},
  \bibinfo {author} {\bibfnamefont {Ashvin}\ \bibnamefont {Vishwanath}}, \ and\
  \bibinfo {author} {\bibfnamefont {Xiangang}\ \bibnamefont {Wan}},\ }\bibfield
   {title} {\enquote {\bibinfo {title} {Comprehensive search for topological
  materials using symmetry indicators},}\ }\href {\doibase
  10.1038/s41586-019-0937-5} {\bibfield  {journal} {\bibinfo  {journal}
  {Nature}\ }\textbf {\bibinfo {volume} {566}},\ \bibinfo {pages} {486--489}
  (\bibinfo {year} {2019})}\BibitemShut {NoStop}%
\bibitem [{\citenamefont {Xu}\ \emph {et~al.}(2020)\citenamefont {Xu} \emph
  {et~al.}}]{MTQCdata2020Xu}%
  \BibitemOpen
  \bibfield  {author} {\bibinfo {author} {\bibfnamefont {Yuanfeng}\
  \bibnamefont {Xu}} \emph {et~al.},\ }\bibfield  {title} {\enquote {\bibinfo
  {title} {High-throughput calculations of magnetic topological materials},}\
  }\href {\doibase 10.1038/s41586-020-2837-0} {\bibfield  {journal} {\bibinfo
  {journal} {Nature}\ }\textbf {\bibinfo {volume} {586}},\ \bibinfo {pages}
  {702--707} (\bibinfo {year} {2020})}\BibitemShut {NoStop}%
\bibitem [{\citenamefont {Gao}\ \emph {et~al.}(2020)\citenamefont {Gao},
  \citenamefont {Peng}, \citenamefont {Wang}, \citenamefont {Fang},\ and\
  \citenamefont {Weng}}]{GaojcNSR}%
  \BibitemOpen
  \bibfield  {author} {\bibinfo {author} {\bibfnamefont {Jiacheng}\
  \bibnamefont {Gao}}, \bibinfo {author} {\bibfnamefont {Shiyu}\ \bibnamefont
  {Peng}}, \bibinfo {author} {\bibfnamefont {Zhijun}\ \bibnamefont {Wang}},
  \bibinfo {author} {\bibfnamefont {Chen}\ \bibnamefont {Fang}}, \ and\
  \bibinfo {author} {\bibfnamefont {Hongming}\ \bibnamefont {Weng}},\
  }\bibfield  {title} {\enquote {\bibinfo {title} {{Electronic structures and
  topological properties in nickelates
  $\mathrm{Ln}_{n+1}\mathrm{Ni}_{n}O_{2n+2}$}},}\ }\href
  {https://doi.org/10.1093/nsr/nwaa218} {\bibfield  {journal} {\bibinfo
  {journal} {Nat. Sci. Rev.}\ }\textbf {\bibinfo {volume} {8}} (\bibinfo {year}
  {2020})}\BibitemShut {NoStop}%
\bibitem [{\citenamefont {Shi}\ \emph {et~al.}(2021)\citenamefont {Shi} \emph
  {et~al.}}]{wzjCDWTSM}%
  \BibitemOpen
  \bibfield  {author} {\bibinfo {author} {\bibfnamefont {Wujun}\ \bibnamefont
  {Shi}} \emph {et~al.},\ }\bibfield  {title} {\enquote {\bibinfo {title} {A
  charge-density-wave topological semimetal},}\ }\href {\doibase
  10.1038/s41567-020-01104-z} {\bibfield  {journal} {\bibinfo  {journal} {Nat.
  Phys.}\ }\textbf {\bibinfo {volume} {17}},\ \bibinfo {pages} {381--387}
  (\bibinfo {year} {2021})}\BibitemShut {NoStop}%
\bibitem [{\citenamefont {Nie}\ \emph {et~al.}(2020)\citenamefont {Nie},
  \citenamefont {Sun}, \citenamefont {Prinz}, \citenamefont {Wang},
  \citenamefont {Weng}, \citenamefont {Fang},\ and\ \citenamefont
  {Dai}}]{PhysRevLett.124.076403QAHEuB6}%
  \BibitemOpen
  \bibfield  {author} {\bibinfo {author} {\bibfnamefont {Simin}\ \bibnamefont
  {Nie}}, \bibinfo {author} {\bibfnamefont {Yan}\ \bibnamefont {Sun}}, \bibinfo
  {author} {\bibfnamefont {Fritz~B.}\ \bibnamefont {Prinz}}, \bibinfo {author}
  {\bibfnamefont {Zhijun}\ \bibnamefont {Wang}}, \bibinfo {author}
  {\bibfnamefont {Hongming}\ \bibnamefont {Weng}}, \bibinfo {author}
  {\bibfnamefont {Zhong}\ \bibnamefont {Fang}}, \ and\ \bibinfo {author}
  {\bibfnamefont {Xi}~\bibnamefont {Dai}},\ }\bibfield  {title} {\enquote
  {\bibinfo {title} {Magnetic semimetals and quantized anomalous hall effect in
  ${\mathrm{eub}}_{6}$},}\ }\href {\doibase 10.1103/PhysRevLett.124.076403}
  {\bibfield  {journal} {\bibinfo  {journal} {Phys. Rev. Lett.}\ }\textbf
  {\bibinfo {volume} {124}},\ \bibinfo {pages} {076403} (\bibinfo {year}
  {2020})}\BibitemShut {NoStop}%
\bibitem [{\citenamefont {Guo}\ \emph {et~al.}(2021)\citenamefont {Guo},
  \citenamefont {Yan}, \citenamefont {Sheng}, \citenamefont {Nie},
  \citenamefont {Shi},\ and\ \citenamefont {Wang}}]{PhysRevB.103.115145QSH235}%
  \BibitemOpen
  \bibfield  {author} {\bibinfo {author} {\bibfnamefont {Zhaopeng}\
  \bibnamefont {Guo}}, \bibinfo {author} {\bibfnamefont {Dayu}\ \bibnamefont
  {Yan}}, \bibinfo {author} {\bibfnamefont {Haohao}\ \bibnamefont {Sheng}},
  \bibinfo {author} {\bibfnamefont {Simin}\ \bibnamefont {Nie}}, \bibinfo
  {author} {\bibfnamefont {Youguo}\ \bibnamefont {Shi}}, \ and\ \bibinfo
  {author} {\bibfnamefont {Zhijun}\ \bibnamefont {Wang}},\ }\bibfield  {title}
  {\enquote {\bibinfo {title} {Quantum spin hall effect in
  $\mathrm{Ta}_{2}{M}_{3}\mathrm{Te}_{5}$ $({M}=\mathrm{Pd}, \mathrm{Ni})$},}\
  }\href {\doibase 10.1103/PhysRevB.103.115145} {\bibfield  {journal} {\bibinfo
   {journal} {Phys. Rev. B}\ }\textbf {\bibinfo {volume} {103}},\ \bibinfo
  {pages} {115145} (\bibinfo {year} {2021})}\BibitemShut {NoStop}%
\bibitem [{\citenamefont {Guo}\ \emph {et~al.}(2022)\citenamefont {Guo},
  \citenamefont {Deng}, \citenamefont {Xie},\ and\ \citenamefont
  {Wang}}]{GuozpQTI2022}%
  \BibitemOpen
  \bibfield  {author} {\bibinfo {author} {\bibfnamefont {Z.}~\bibnamefont
  {Guo}}, \bibinfo {author} {\bibfnamefont {J.}~\bibnamefont {Deng}}, \bibinfo
  {author} {\bibfnamefont {Y.}~\bibnamefont {Xie}}, \ and\ \bibinfo {author}
  {\bibfnamefont {Z.}~\bibnamefont {Wang}},\ }\bibfield  {title} {\enquote
  {\bibinfo {title} {Quadrupole topological insulators in
  $\mathrm{Ta}_{2}{M}_{3}\mathrm{Te}_{5}$ $({M}=\mathrm{Ni}, \mathrm{Pd})$
  monolayers},}\ }\href {\doibase 10.1038/s41535-022-00498-8} {\bibfield
  {journal} {\bibinfo  {journal} {NPJ Quantum Mater.}\ }\textbf {\bibinfo
  {volume} {7}},\ \bibinfo {pages} {87} (\bibinfo {year} {2022})}\BibitemShut
  {NoStop}%
\bibitem [{\citenamefont {Nie}\ \emph {et~al.}(2018)\citenamefont {Nie},
  \citenamefont {Xing}, \citenamefont {Jin}, \citenamefont {Xie}, \citenamefont
  {Wang},\ and\ \citenamefont {Prinz}}]{PhysRevB.98.125143TaSe3}%
  \BibitemOpen
  \bibfield  {author} {\bibinfo {author} {\bibfnamefont {Simin}\ \bibnamefont
  {Nie}}, \bibinfo {author} {\bibfnamefont {Lingyi}\ \bibnamefont {Xing}},
  \bibinfo {author} {\bibfnamefont {Rongying}\ \bibnamefont {Jin}}, \bibinfo
  {author} {\bibfnamefont {Weiwei}\ \bibnamefont {Xie}}, \bibinfo {author}
  {\bibfnamefont {Zhijun}\ \bibnamefont {Wang}}, \ and\ \bibinfo {author}
  {\bibfnamefont {Fritz~B.}\ \bibnamefont {Prinz}},\ }\bibfield  {title}
  {\enquote {\bibinfo {title} {Topological phases in the $\mathrm{TaSe_{3}}$
  compound},}\ }\href {\doibase 10.1103/PhysRevB.98.125143} {\bibfield
  {journal} {\bibinfo  {journal} {Phys. Rev. B}\ }\textbf {\bibinfo {volume}
  {98}},\ \bibinfo {pages} {125143} (\bibinfo {year} {2018})}\BibitemShut
  {NoStop}%
\bibitem [{\citenamefont {Nie}\ \emph {et~al.}(2021{\natexlab{a}})\citenamefont
  {Nie}, \citenamefont {Qian}, \citenamefont {Gao}, \citenamefont {Fang},
  \citenamefont {Weng},\ and\ \citenamefont
  {Wang}}]{PhysRevB.103.205133electrides}%
  \BibitemOpen
  \bibfield  {author} {\bibinfo {author} {\bibfnamefont {Simin}\ \bibnamefont
  {Nie}}, \bibinfo {author} {\bibfnamefont {Yuting}\ \bibnamefont {Qian}},
  \bibinfo {author} {\bibfnamefont {Jiacheng}\ \bibnamefont {Gao}}, \bibinfo
  {author} {\bibfnamefont {Zhong}\ \bibnamefont {Fang}}, \bibinfo {author}
  {\bibfnamefont {Hongming}\ \bibnamefont {Weng}}, \ and\ \bibinfo {author}
  {\bibfnamefont {Zhijun}\ \bibnamefont {Wang}},\ }\bibfield  {title} {\enquote
  {\bibinfo {title} {Application of topological quantum chemistry in
  electrides},}\ }\href {\doibase 10.1103/PhysRevB.103.205133} {\bibfield
  {journal} {\bibinfo  {journal} {Phys. Rev. B}\ }\textbf {\bibinfo {volume}
  {103}},\ \bibinfo {pages} {205133} (\bibinfo {year}
  {2021}{\natexlab{a}})}\BibitemShut {NoStop}%
\bibitem [{\citenamefont {Nie}\ \emph {et~al.}(2021{\natexlab{b}})\citenamefont
  {Nie}, \citenamefont {Bernevig},\ and\ \citenamefont {Wang}}]{niesimin_prr}%
  \BibitemOpen
  \bibfield  {author} {\bibinfo {author} {\bibfnamefont {Simin}\ \bibnamefont
  {Nie}}, \bibinfo {author} {\bibfnamefont {B.~Andrei}\ \bibnamefont
  {Bernevig}}, \ and\ \bibinfo {author} {\bibfnamefont {Zhijun}\ \bibnamefont
  {Wang}},\ }\bibfield  {title} {\enquote {\bibinfo {title} {Sixfold
  excitations in electrides},}\ }\href {\doibase
  10.1103/PhysRevResearch.3.L012028} {\bibfield  {journal} {\bibinfo  {journal}
  {Phys. Rev. Research}\ }\textbf {\bibinfo {volume} {3}},\ \bibinfo {pages}
  {L012028} (\bibinfo {year} {2021}{\natexlab{b}})}\BibitemShut {NoStop}%
\bibitem [{\citenamefont {Gao}\ \emph {et~al.}(2022{\natexlab{b}})\citenamefont
  {Gao} \emph {et~al.}}]{GaojcUnconvMat}%
  \BibitemOpen
  \bibfield  {author} {\bibinfo {author} {\bibfnamefont {Jiacheng}\
  \bibnamefont {Gao}} \emph {et~al.},\ }\bibfield  {title} {\enquote {\bibinfo
  {title} {Unconventional materials: the mismatch between electronic charge
  centers and atomic positions},}\ }\href {\doibase
  https://doi.org/10.1016/j.scib.2021.12.025} {\bibfield  {journal} {\bibinfo
  {journal} {Sci. Bull.}\ }\textbf {\bibinfo {volume} {67}},\ \bibinfo {pages}
  {598--608} (\bibinfo {year} {2022}{\natexlab{b}})}\BibitemShut {NoStop}%
\bibitem [{\citenamefont {{Xu}}\ \emph {et~al.}()\citenamefont {{Xu}} \emph
  {et~al.}}]{xu2021threedimensional}%
  \BibitemOpen
  \bibfield  {author} {\bibinfo {author} {\bibfnamefont {Yuanfeng}\
  \bibnamefont {{Xu}}} \emph {et~al.},\ }\href@noop {} {\enquote {\bibinfo
  {title} {{Three-Dimensional Real Space Invariants, Obstructed Atomic
  Insulators and A New Principle for Active Catalytic Sites}},}\ }\Eprint
  {http://arxiv.org/abs/2111.02433} {arXiv:2111.02433} \BibitemShut {NoStop}%
\bibitem [{\citenamefont {Li}(2022)}]{guoweili}%
  \BibitemOpen
  \bibfield  {author} {\bibinfo {author} {\bibfnamefont {Guowei et~al.}\
  \bibnamefont {Li}},\ }\bibfield  {title} {\enquote {\bibinfo {title}
  {Obstructed surface states as the descriptor for predicting catalytic active
  sites in inorganic crystalline materials},}\ }\href {\doibase
  https://doi.org/10.1002/adma.202201328} {\bibfield  {journal} {\bibinfo
  {journal} {Adv. Mater.}\ }\textbf {\bibinfo {volume} {34}},\ \bibinfo {pages}
  {2201328} (\bibinfo {year} {2022})}\BibitemShut {NoStop}%
\bibitem [{\citenamefont {Sivula}\ and\ \citenamefont {van~de
  Krol}(2016)}]{interface1_usage}%
  \BibitemOpen
  \bibfield  {author} {\bibinfo {author} {\bibfnamefont {Kevin}\ \bibnamefont
  {Sivula}}\ and\ \bibinfo {author} {\bibfnamefont {Roel}\ \bibnamefont {van~de
  Krol}},\ }\bibfield  {title} {\enquote {\bibinfo {title} {Semiconducting
  materials for photoelectrochemical energy conversion},}\ }\href {\doibase
  10.1038/natrevmats.2015.10} {\bibfield  {journal} {\bibinfo  {journal} {Nat.
  Rev. Mater.}\ }\textbf {\bibinfo {volume} {1}},\ \bibinfo {pages} {15010}
  (\bibinfo {year} {2016})}\BibitemShut {NoStop}%
\bibitem [{\citenamefont {Wu}\ \emph {et~al.}(2022)\citenamefont {Wu} \emph
  {et~al.}}]{ferroelecNature}%
  \BibitemOpen
  \bibfield  {author} {\bibinfo {author} {\bibfnamefont {Heng}\ \bibnamefont
  {Wu}} \emph {et~al.},\ }\bibfield  {title} {\enquote {\bibinfo {title} {The
  field-free josephson diode in a van der waals heterostructure},}\ }\href
  {\doibase 10.1038/s41586-022-04504-8} {\bibfield  {journal} {\bibinfo
  {journal} {Nature}\ }\textbf {\bibinfo {volume} {604}},\ \bibinfo {pages}
  {653--656} (\bibinfo {year} {2022})}\BibitemShut {NoStop}%
\bibitem [{\citenamefont {Reyren}\ \emph {et~al.}(2007)\citenamefont {Reyren}
  \emph {et~al.}}]{supr-con1_usage}%
  \BibitemOpen
  \bibfield  {author} {\bibinfo {author} {\bibfnamefont {N.}~\bibnamefont
  {Reyren}} \emph {et~al.},\ }\bibfield  {title} {\enquote {\bibinfo {title}
  {Superconducting interfaces between insulating oxides},}\ }\href {\doibase
  10.1126/science.1146006} {\bibfield  {journal} {\bibinfo  {journal}
  {Science}\ }\textbf {\bibinfo {volume} {317}},\ \bibinfo {pages} {1196--1199}
  (\bibinfo {year} {2007})}\BibitemShut {NoStop}%
\bibitem [{\citenamefont {Ohtomo}\ and\ \citenamefont
  {Hwang}(2004)}]{supr-con2_usage}%
  \BibitemOpen
  \bibfield  {author} {\bibinfo {author} {\bibfnamefont {A.}~\bibnamefont
  {Ohtomo}}\ and\ \bibinfo {author} {\bibfnamefont {H.~Y.}\ \bibnamefont
  {Hwang}},\ }\bibfield  {title} {\enquote {\bibinfo {title} {A high-mobility
  electron gas at the $\mathrm{LaAlO_3/SrTiO_3}$ heterointerface},}\ }\href
  {\doibase 10.1038/nature02308} {\bibfield  {journal} {\bibinfo  {journal}
  {Nature}\ }\textbf {\bibinfo {volume} {427}},\ \bibinfo {pages} {423--426}
  (\bibinfo {year} {2004})}\BibitemShut {NoStop}%
\bibitem [{\citenamefont {Benalcazar}\ \emph
  {et~al.}(2017{\natexlab{a}})\citenamefont {Benalcazar}, \citenamefont
  {Bernevig},\ and\ \citenamefont {Hughes}}]{multipoleScience}%
  \BibitemOpen
  \bibfield  {author} {\bibinfo {author} {\bibfnamefont {Wladimir~A.}\
  \bibnamefont {Benalcazar}}, \bibinfo {author} {\bibfnamefont {B.~Andrei}\
  \bibnamefont {Bernevig}}, \ and\ \bibinfo {author} {\bibfnamefont
  {Taylor~L.}\ \bibnamefont {Hughes}},\ }\bibfield  {title} {\enquote {\bibinfo
  {title} {Quantized electric multipole insulators},}\ }\href {\doibase
  10.1126/science.aah6442} {\bibfield  {journal} {\bibinfo  {journal}
  {Science}\ }\textbf {\bibinfo {volume} {357}},\ \bibinfo {pages} {61--66}
  (\bibinfo {year} {2017}{\natexlab{a}})}\BibitemShut {NoStop}%
\bibitem [{\citenamefont {Benalcazar}\ \emph
  {et~al.}(2017{\natexlab{b}})\citenamefont {Benalcazar}, \citenamefont
  {Bernevig},\ and\ \citenamefont {Hughes}}]{PhysRevB.96.245115multipole}%
  \BibitemOpen
  \bibfield  {author} {\bibinfo {author} {\bibfnamefont {Wladimir~A.}\
  \bibnamefont {Benalcazar}}, \bibinfo {author} {\bibfnamefont {B.~Andrei}\
  \bibnamefont {Bernevig}}, \ and\ \bibinfo {author} {\bibfnamefont
  {Taylor~L.}\ \bibnamefont {Hughes}},\ }\bibfield  {title} {\enquote {\bibinfo
  {title} {Electric multipole moments, topological multipole moment pumping,
  and chiral hinge states in crystalline insulators},}\ }\href {\doibase
  10.1103/PhysRevB.96.245115} {\bibfield  {journal} {\bibinfo  {journal} {Phys.
  Rev. B}\ }\textbf {\bibinfo {volume} {96}},\ \bibinfo {pages} {245115}
  (\bibinfo {year} {2017}{\natexlab{b}})}\BibitemShut {NoStop}%
\bibitem [{\citenamefont {Schindler}\ \emph {et~al.}(2019)\citenamefont
  {Schindler}, \citenamefont {Brzezi\ifmmode~\acute{n}\else \'{n}\fi{}ska},
  \citenamefont {Benalcazar}, \citenamefont {Iraola}, \citenamefont {Bouhon},
  \citenamefont {Tsirkin}, \citenamefont {Vergniory},\ and\ \citenamefont
  {Neupert}}]{PhysRevResearch.1.033074fractionalCorner}%
  \BibitemOpen
  \bibfield  {author} {\bibinfo {author} {\bibfnamefont {Frank}\ \bibnamefont
  {Schindler}}, \bibinfo {author} {\bibfnamefont {Marta}\ \bibnamefont
  {Brzezi\ifmmode~\acute{n}\else \'{n}\fi{}ska}}, \bibinfo {author}
  {\bibfnamefont {Wladimir~A.}\ \bibnamefont {Benalcazar}}, \bibinfo {author}
  {\bibfnamefont {Mikel}\ \bibnamefont {Iraola}}, \bibinfo {author}
  {\bibfnamefont {Adrien}\ \bibnamefont {Bouhon}}, \bibinfo {author}
  {\bibfnamefont {Stepan~S.}\ \bibnamefont {Tsirkin}}, \bibinfo {author}
  {\bibfnamefont {Maia~G.}\ \bibnamefont {Vergniory}}, \ and\ \bibinfo {author}
  {\bibfnamefont {Titus}\ \bibnamefont {Neupert}},\ }\bibfield  {title}
  {\enquote {\bibinfo {title} {Fractional corner charges in spin-orbit coupled
  crystals},}\ }\href {\doibase 10.1103/PhysRevResearch.1.033074} {\bibfield
  {journal} {\bibinfo  {journal} {Phys. Rev. Research}\ }\textbf {\bibinfo
  {volume} {1}},\ \bibinfo {pages} {033074} (\bibinfo {year}
  {2019})}\BibitemShut {NoStop}%
\bibitem [{\citenamefont {Xu}\ \emph {et~al.}()\citenamefont {Xu} \emph
  {et~al.}}]{xu2021fillingenforced}%
  \BibitemOpen
  \bibfield  {author} {\bibinfo {author} {\bibfnamefont {Yuanfeng}\
  \bibnamefont {Xu}} \emph {et~al.},\ }\href@noop {} {\enquote {\bibinfo
  {title} {{Filling-Enforced Obstructed Atomic Insulators}},}\ }\Eprint
  {http://arxiv.org/abs/2106.10276} {arXiv:2106.10276} \BibitemShut {NoStop}%
\bibitem [{\citenamefont {Zhang}\ \emph {et~al.}(2023)\citenamefont {Zhang},
  \citenamefont {Sheng}, \citenamefont {Deng}, \citenamefont {Yang},\ and\
  \citenamefont {Wang}}]{zhang2023unconventional}%
  \BibitemOpen
  \bibfield  {author} {\bibinfo {author} {\bibfnamefont {Ruihan}\ \bibnamefont
  {Zhang}}, \bibinfo {author} {\bibfnamefont {Haohao}\ \bibnamefont {Sheng}},
  \bibinfo {author} {\bibfnamefont {Junze}\ \bibnamefont {Deng}}, \bibinfo
  {author} {\bibfnamefont {Zhilong}\ \bibnamefont {Yang}}, \ and\ \bibinfo
  {author} {\bibfnamefont {Zhijun}\ \bibnamefont {Wang}},\ }\href@noop {}
  {\enquote {\bibinfo {title} {Unconventional phonon spectra and obstructed
  edge phonon modes},}\ } (\bibinfo {year} {2023}),\ \Eprint
  {http://arxiv.org/abs/2305.09453} {arXiv:2305.09453 [cond-mat.mtrl-sci]}
  \BibitemShut {NoStop}%
\bibitem [{\citenamefont {compute phonon-mode irreducible representations with
  interface to~the Quantum ESPRESSO~package}()}]{ir2ph}%
  \BibitemOpen
  \bibfield  {author} {\bibinfo {author} {\bibfnamefont {To}~\bibnamefont
  {compute phonon-mode irreducible representations with interface to~the
  Quantum ESPRESSO~package}},\ }\href {https://github.com/zjwang11/ir2pw.}
  {\bibinfo  {journal} {https://github.com/zjwang11/ir2pw.}\ }\BibitemShut
  {NoStop}%
\bibitem [{\citenamefont {eBR/aBR decomposition: a general workflow to get
  unconventional~band structures}()}]{unconvmat}%
  \BibitemOpen
\bibfield  {journal} {  }\bibfield  {author} {\bibinfo {author} {\bibnamefont
  {eBR/aBR decomposition: a general workflow to get unconventional~band
  structures}},\ }\href {http://tm.iphy.ac.cn/UnconvMat.html} {\bibinfo
  {journal} {http://tm.iphy.ac.cn/UnconvMat.html}\ }\BibitemShut {NoStop}%
\bibitem [{\citenamefont {Gao}\ \emph {et~al.}(2021)\citenamefont {Gao},
  \citenamefont {Wu}, \citenamefont {Persson},\ and\ \citenamefont
  {Wang}}]{GaojcIRVSP}%
  \BibitemOpen
\bibfield  {journal} {  }\bibfield  {author} {\bibinfo {author} {\bibfnamefont
  {Jiacheng}\ \bibnamefont {Gao}}, \bibinfo {author} {\bibfnamefont
  {Quansheng}\ \bibnamefont {Wu}}, \bibinfo {author} {\bibfnamefont {Clas}\
  \bibnamefont {Persson}}, \ and\ \bibinfo {author} {\bibfnamefont {Zhijun}\
  \bibnamefont {Wang}},\ }\bibfield  {title} {\enquote {\bibinfo {title}
  {Irvsp: To obtain irreducible representations of electronic states in the
  {VASP}},}\ }\href {\doibase 10.1016/j.cpc.2020.107760} {\bibfield  {journal}
  {\bibinfo  {journal} {Comput. Phys. Commun.}\ }\textbf {\bibinfo {volume}
  {261}},\ \bibinfo {pages} {107760} (\bibinfo {year} {2021})}\BibitemShut
  {NoStop}%
\bibitem [{\citenamefont {Zak}(1989)}]{zak}%
  \BibitemOpen
  \bibfield  {author} {\bibinfo {author} {\bibfnamefont {J.}~\bibnamefont
  {Zak}},\ }\bibfield  {title} {\enquote {\bibinfo {title} {Berry's phase for
  energy bands in solids},}\ }\href {\doibase 10.1103/PhysRevLett.62.2747}
  {\bibfield  {journal} {\bibinfo  {journal} {Phys. Rev. Lett.}\ }\textbf
  {\bibinfo {volume} {62}},\ \bibinfo {pages} {2747--2750} (\bibinfo {year}
  {1989})}\BibitemShut {NoStop}%
\bibitem [{\citenamefont {Rhim}\ \emph {et~al.}(2017)\citenamefont {Rhim},
  \citenamefont {Behrends},\ and\ \citenamefont {Bardarson}}]{zakphase}%
  \BibitemOpen
  \bibfield  {author} {\bibinfo {author} {\bibfnamefont {Jun-Won}\ \bibnamefont
  {Rhim}}, \bibinfo {author} {\bibfnamefont {Jan}\ \bibnamefont {Behrends}}, \
  and\ \bibinfo {author} {\bibfnamefont {Jens~H.}\ \bibnamefont {Bardarson}},\
  }\bibfield  {title} {\enquote {\bibinfo {title} {Bulk-boundary correspondence
  from the intercellular zak phase},}\ }\href {\doibase
  10.1103/PhysRevB.95.035421} {\bibfield  {journal} {\bibinfo  {journal} {Phys.
  Rev. B}\ }\textbf {\bibinfo {volume} {95}},\ \bibinfo {pages} {035421}
  (\bibinfo {year} {2017})}\BibitemShut {NoStop}%
\bibitem [{\citenamefont {Kraut}\ and\ \citenamefont {von Baltz}(1979)}]{sc1}%
  \BibitemOpen
  \bibfield  {author} {\bibinfo {author} {\bibfnamefont {Wolfgang}\
  \bibnamefont {Kraut}}\ and\ \bibinfo {author} {\bibfnamefont {Ralph}\
  \bibnamefont {von Baltz}},\ }\bibfield  {title} {\enquote {\bibinfo {title}
  {Anomalous bulk photovoltaic effect in ferroelectrics: A quadratic response
  theory},}\ }\href {\doibase 10.1103/PhysRevB.19.1548} {\bibfield  {journal}
  {\bibinfo  {journal} {Phys. Rev. B}\ }\textbf {\bibinfo {volume} {19}},\
  \bibinfo {pages} {1548--1554} (\bibinfo {year} {1979})}\BibitemShut {NoStop}%
\bibitem [{\citenamefont {Belinicher}\ and\ \citenamefont
  {Sturman}(1980)}]{sc2}%
  \BibitemOpen
  \bibfield  {author} {\bibinfo {author} {\bibfnamefont {V~I}\ \bibnamefont
  {Belinicher}}\ and\ \bibinfo {author} {\bibfnamefont {B~I}\ \bibnamefont
  {Sturman}},\ }\bibfield  {title} {\enquote {\bibinfo {title} {The
  photogalvanic effect in media lacking a center of symmetry},}\ }\href
  {\doibase 10.1070/pu1980v023n03abeh004703} {\bibfield  {journal} {\bibinfo
  {journal} {Sov. Phys. Usp.}\ }\textbf {\bibinfo {volume} {23}},\ \bibinfo
  {pages} {199--223} (\bibinfo {year} {1980})}\BibitemShut {NoStop}%
\bibitem [{\citenamefont {Sipe}\ and\ \citenamefont {Shkrebtii}(2000)}]{sc3}%
  \BibitemOpen
  \bibfield  {author} {\bibinfo {author} {\bibfnamefont {J.~E.}\ \bibnamefont
  {Sipe}}\ and\ \bibinfo {author} {\bibfnamefont {A.~I.}\ \bibnamefont
  {Shkrebtii}},\ }\bibfield  {title} {\enquote {\bibinfo {title} {Second-order
  optical response in semiconductors},}\ }\href {\doibase
  10.1103/PhysRevB.61.5337} {\bibfield  {journal} {\bibinfo  {journal} {Phys.
  Rev. B}\ }\textbf {\bibinfo {volume} {61}},\ \bibinfo {pages} {5337--5352}
  (\bibinfo {year} {2000})}\BibitemShut {NoStop}%
\bibitem [{\citenamefont {Young}\ and\ \citenamefont {Rappe}(2012)}]{m1}%
  \BibitemOpen
  \bibfield  {author} {\bibinfo {author} {\bibfnamefont {Steve~M.}\
  \bibnamefont {Young}}\ and\ \bibinfo {author} {\bibfnamefont {Andrew~M.}\
  \bibnamefont {Rappe}},\ }\bibfield  {title} {\enquote {\bibinfo {title}
  {First principles calculation of the shift current photovoltaic effect in
  ferroelectrics},}\ }\href {\doibase 10.1103/PhysRevLett.109.116601}
  {\bibfield  {journal} {\bibinfo  {journal} {Phys. Rev. Lett.}\ }\textbf
  {\bibinfo {volume} {109}},\ \bibinfo {pages} {116601} (\bibinfo {year}
  {2012})}\BibitemShut {NoStop}%
\bibitem [{\citenamefont {Zheng}\ \emph {et~al.}(2015)\citenamefont {Zheng},
  \citenamefont {Takenaka}, \citenamefont {Wang}, \citenamefont {Koocher},\
  and\ \citenamefont {Rappe}}]{m2}%
  \BibitemOpen
  \bibfield  {author} {\bibinfo {author} {\bibfnamefont {Fan}\ \bibnamefont
  {Zheng}}, \bibinfo {author} {\bibfnamefont {Hiroyuki}\ \bibnamefont
  {Takenaka}}, \bibinfo {author} {\bibfnamefont {Fenggong}\ \bibnamefont
  {Wang}}, \bibinfo {author} {\bibfnamefont {Nathan~Z.}\ \bibnamefont
  {Koocher}}, \ and\ \bibinfo {author} {\bibfnamefont {Andrew~M.}\ \bibnamefont
  {Rappe}},\ }\bibfield  {title} {\enquote {\bibinfo {title} {First-principles
  calculation of the bulk photovoltaic effect in
  $\mathrm{CH}_{3}\mathrm{NH}{3}\mathrm{PbI}_{3}$ and
  $\mathrm{CH}_{3}\mathrm{NH}_{3}\mathrm{PbI}_{3-x}\mathrm{Cl}_{x}$},}\ }\href
  {\doibase 10.1021/jz502109e} {\bibfield  {journal} {\bibinfo  {journal} {J.
  Phys. Chem. Lett.}\ }\textbf {\bibinfo {volume} {6}},\ \bibinfo {pages}
  {31--37} (\bibinfo {year} {2015})}\BibitemShut {NoStop}%
\bibitem [{\citenamefont {Brehm}\ \emph {et~al.}(2014)\citenamefont {Brehm},
  \citenamefont {Young}, \citenamefont {Zheng},\ and\ \citenamefont
  {Rappe}}]{m3}%
  \BibitemOpen
  \bibfield  {author} {\bibinfo {author} {\bibfnamefont {John~A.}\ \bibnamefont
  {Brehm}}, \bibinfo {author} {\bibfnamefont {Steve~M.}\ \bibnamefont {Young}},
  \bibinfo {author} {\bibfnamefont {Fan}\ \bibnamefont {Zheng}}, \ and\
  \bibinfo {author} {\bibfnamefont {Andrew~M.}\ \bibnamefont {Rappe}},\
  }\bibfield  {title} {\enquote {\bibinfo {title} {First-principles calculation
  of the bulk photovoltaic effect in the polar compounds $\mathrm{Li}
  \mathrm{As} \mathrm{S}_{2}, \mathrm{Li} \mathrm{As} \mathrm{Se}_{2},$ and
  $\mathrm{Na} \mathrm{As} \mathrm{Se}_{2}$},}\ }\href {\doibase
  10.1063/1.4901433} {\bibfield  {journal} {\bibinfo  {journal} {J. Chem.
  Phys.}\ }\textbf {\bibinfo {volume} {141}},\ \bibinfo {pages} {204704}
  (\bibinfo {year} {2014})}\BibitemShut {NoStop}%
\bibitem [{\citenamefont {von Hippel}(1950)}]{m4}%
  \BibitemOpen
  \bibfield  {author} {\bibinfo {author} {\bibfnamefont {A.}~\bibnamefont {von
  Hippel}},\ }\bibfield  {title} {\enquote {\bibinfo {title} {Ferroelectricity,
  domain structure, and phase transitions of barium titanate},}\ }\href
  {\doibase 10.1103/RevModPhys.22.221} {\bibfield  {journal} {\bibinfo
  {journal} {Rev. Mod. Phys.}\ }\textbf {\bibinfo {volume} {22}},\ \bibinfo
  {pages} {221--237} (\bibinfo {year} {1950})}\BibitemShut {NoStop}%
\bibitem [{\citenamefont {Shieh}\ \emph {et~al.}(2009)\citenamefont {Shieh},
  \citenamefont {Yeh}, \citenamefont {Shu},\ and\ \citenamefont {Yen}}]{m5}%
  \BibitemOpen
  \bibfield  {author} {\bibinfo {author} {\bibfnamefont {J.}~\bibnamefont
  {Shieh}}, \bibinfo {author} {\bibfnamefont {J.H.}\ \bibnamefont {Yeh}},
  \bibinfo {author} {\bibfnamefont {Y.C.}\ \bibnamefont {Shu}}, \ and\ \bibinfo
  {author} {\bibfnamefont {J.H.}\ \bibnamefont {Yen}},\ }\bibfield  {title}
  {\enquote {\bibinfo {title} {Hysteresis behaviors of barium titanate single
  crystals based on the operation of multiple $90^{\circ}$ switching
  systems},}\ }\href {\doibase https://doi.org/10.1016/j.mseb.2008.11.046}
  {\bibfield  {journal} {\bibinfo  {journal} {Mater. Sci. Eng. B}\ }\textbf
  {\bibinfo {volume} {161}},\ \bibinfo {pages} {50--54} (\bibinfo {year}
  {2009})}\BibitemShut {NoStop}%
\bibitem [{\citenamefont {Liu}\ \emph {et~al.}(2017)\citenamefont {Liu},
  \citenamefont {Zheng},\ and\ \citenamefont {Rappe}}]{m6}%
  \BibitemOpen
  \bibfield  {author} {\bibinfo {author} {\bibfnamefont {Shi}\ \bibnamefont
  {Liu}}, \bibinfo {author} {\bibfnamefont {Fan}\ \bibnamefont {Zheng}}, \ and\
  \bibinfo {author} {\bibfnamefont {Andrew~M.}\ \bibnamefont {Rappe}},\
  }\bibfield  {title} {\enquote {\bibinfo {title} {Giant bulk photovoltaic
  effect in vinylene-linked hybrid heterocyclic polymer},}\ }\href {\doibase
  10.1021/acs.jpcc.7b00374} {\bibfield  {journal} {\bibinfo  {journal} {J.
  Phys. Chem. C}\ }\textbf {\bibinfo {volume} {121}},\ \bibinfo {pages}
  {6500--6507} (\bibinfo {year} {2017})}\BibitemShut {NoStop}%
\bibitem [{\citenamefont {C{\^o}t{\'e}}\ \emph {et~al.}(2002)\citenamefont
  {C{\^o}t{\'e}}, \citenamefont {Laman},\ and\ \citenamefont {van Driel}}]{m7}%
  \BibitemOpen
  \bibfield  {author} {\bibinfo {author} {\bibfnamefont {D.}~\bibnamefont
  {C{\^o}t{\'e}}}, \bibinfo {author} {\bibfnamefont {N.}~\bibnamefont {Laman}},
  \ and\ \bibinfo {author} {\bibfnamefont {H.~M.}\ \bibnamefont {van Driel}},\
  }\bibfield  {title} {\enquote {\bibinfo {title} {Rectification and shift
  currents in gaas},}\ }\href {\doibase 10.1063/1.1436530} {\bibfield
  {journal} {\bibinfo  {journal} {Appl. Phys. Lett.}\ }\textbf {\bibinfo
  {volume} {80}},\ \bibinfo {pages} {905--907} (\bibinfo {year}
  {2002})}\BibitemShut {NoStop}%
\bibitem [{\citenamefont {Sotome}\ \emph {et~al.}(2019)\citenamefont {Sotome}
  \emph {et~al.}}]{m8}%
  \BibitemOpen
  \bibfield  {author} {\bibinfo {author} {\bibfnamefont {M.}~\bibnamefont
  {Sotome}} \emph {et~al.},\ }\bibfield  {title} {\enquote {\bibinfo {title}
  {Spectral dynamics of shift current in ferroelectric semiconductor sbsi},}\
  }\href {\doibase 10.1073/pnas.1802427116} {\bibfield  {journal} {\bibinfo
  {journal} {Proc. Natl Acad. Sci.}\ }\textbf {\bibinfo {volume} {116}},\
  \bibinfo {pages} {1929--1933} (\bibinfo {year} {2019})}\BibitemShut {NoStop}%
\bibitem [{\citenamefont {Tan}\ \emph {et~al.}(2016)\citenamefont {Tan},
  \citenamefont {Zheng}, \citenamefont {Young}, \citenamefont {Wang},
  \citenamefont {Liu},\ and\ \citenamefont {Rappe}}]{m9}%
  \BibitemOpen
  \bibfield  {author} {\bibinfo {author} {\bibfnamefont {Liang~Z}\ \bibnamefont
  {Tan}}, \bibinfo {author} {\bibfnamefont {Fan}\ \bibnamefont {Zheng}},
  \bibinfo {author} {\bibfnamefont {Steve~M}\ \bibnamefont {Young}}, \bibinfo
  {author} {\bibfnamefont {Fenggong}\ \bibnamefont {Wang}}, \bibinfo {author}
  {\bibfnamefont {Shi}\ \bibnamefont {Liu}}, \ and\ \bibinfo {author}
  {\bibfnamefont {Andrew~M}\ \bibnamefont {Rappe}},\ }\bibfield  {title}
  {\enquote {\bibinfo {title} {Shift current bulk photovoltaic effect in polar
  materials---hybrid and oxide perovskites and beyond},}\ }\href {\doibase
  10.1038/npjcompumats.2016.26} {\bibfield  {journal} {\bibinfo  {journal} {NPJ
  Comput. Mater.}\ }\textbf {\bibinfo {volume} {2}},\ \bibinfo {pages} {16026}
  (\bibinfo {year} {2016})}\BibitemShut {NoStop}%
\bibitem [{\citenamefont {Liu}\ \emph {et~al.}(2020)\citenamefont {Liu},
  \citenamefont {Qian}, \citenamefont {Fu},\ and\ \citenamefont
  {Wang}}]{weyl_phonon}%
  \BibitemOpen
  \bibfield  {author} {\bibinfo {author} {\bibfnamefont {Qing-Bo}\ \bibnamefont
  {Liu}}, \bibinfo {author} {\bibfnamefont {Yuting}\ \bibnamefont {Qian}},
  \bibinfo {author} {\bibfnamefont {Hua-Hua}\ \bibnamefont {Fu}}, \ and\
  \bibinfo {author} {\bibfnamefont {Zhijun}\ \bibnamefont {Wang}},\ }\bibfield
  {title} {\enquote {\bibinfo {title} {Symmetry-enforced {Weyl} phonons},}\
  }\href {\doibase 10.1038/s41524-020-00358-8} {\bibfield  {journal} {\bibinfo
  {journal} {NPJ Comput. Mater.}\ }\textbf {\bibinfo {volume} {6}},\ \bibinfo
  {pages} {95} (\bibinfo {year} {2020})}\BibitemShut {NoStop}%
\bibitem [{\citenamefont {K\l{}osi\ifmmode~\acute{n}\else \'{n}\fi{}ski}\ \emph
  {et~al.}(2023)\citenamefont {K\l{}osi\ifmmode~\acute{n}\else \'{n}\fi{}ski},
  \citenamefont {Brzezicki}, \citenamefont {Lau}, \citenamefont {Agrapidis},
  \citenamefont {Ole\ifmmode~\acute{s}\else \'{s}\fi{}}, \citenamefont {van
  Wezel},\ and\ \citenamefont {Wohlfeld}}]{chalcogen}%
  \BibitemOpen
  \bibfield  {author} {\bibinfo {author} {\bibfnamefont {Adam}\ \bibnamefont
  {K\l{}osi\ifmmode~\acute{n}\else \'{n}\fi{}ski}}, \bibinfo {author}
  {\bibfnamefont {Wojciech}\ \bibnamefont {Brzezicki}}, \bibinfo {author}
  {\bibfnamefont {Alexander}\ \bibnamefont {Lau}}, \bibinfo {author}
  {\bibfnamefont {Cli\`o~E.}\ \bibnamefont {Agrapidis}}, \bibinfo {author}
  {\bibfnamefont {Andrzej~M.}\ \bibnamefont {Ole\ifmmode~\acute{s}\else
  \'{s}\fi{}}}, \bibinfo {author} {\bibfnamefont {Jasper}\ \bibnamefont {van
  Wezel}}, \ and\ \bibinfo {author} {\bibfnamefont {Krzysztof}\ \bibnamefont
  {Wohlfeld}},\ }\bibfield  {title} {\enquote {\bibinfo {title} {Topology of
  chalcogen chains},}\ }\href {\doibase 10.1103/PhysRevB.107.125123} {\bibfield
   {journal} {\bibinfo  {journal} {Phys. Rev. B}\ }\textbf {\bibinfo {volume}
  {107}},\ \bibinfo {pages} {125123} (\bibinfo {year} {2023})}\BibitemShut
  {NoStop}%
\bibitem [{\citenamefont {Tran}\ and\ \citenamefont
  {Blaha}(2009)}]{PhysRevLett.102mbj}%
  \BibitemOpen
  \bibfield  {author} {\bibinfo {author} {\bibfnamefont {Fabien}\ \bibnamefont
  {Tran}}\ and\ \bibinfo {author} {\bibfnamefont {Peter}\ \bibnamefont
  {Blaha}},\ }\bibfield  {title} {\enquote {\bibinfo {title} {Accurate band
  gaps of semiconductors and insulators with a semilocal exchange-correlation
  potential},}\ }\href {\doibase 10.1103/PhysRevLett.102.226401} {\bibfield
  {journal} {\bibinfo  {journal} {Phys. Rev. Lett.}\ }\textbf {\bibinfo
  {volume} {102}},\ \bibinfo {pages} {226401} (\bibinfo {year}
  {2009})}\BibitemShut {NoStop}%
\bibitem [{\citenamefont {Becke}\ and\ \citenamefont
  {Johnson}(2006)}]{J.Chem.Phys.124mbj}%
  \BibitemOpen
  \bibfield  {author} {\bibinfo {author} {\bibfnamefont {Axel~D.}\ \bibnamefont
  {Becke}}\ and\ \bibinfo {author} {\bibfnamefont {Erin~R.}\ \bibnamefont
  {Johnson}},\ }\bibfield  {title} {\enquote {\bibinfo {title} {A simple
  effective potential for exchange},}\ }\href {\doibase 10.1063/1.2213970}
  {\bibfield  {journal} {\bibinfo  {journal} {J. Chem. Phys.}\ }\textbf
  {\bibinfo {volume} {124}},\ \bibinfo {pages} {221101} (\bibinfo {year}
  {2006})}\BibitemShut {NoStop}%
\bibitem [{\citenamefont {Giannozzi}\ \emph {et~al.}(2009)\citenamefont
  {Giannozzi} \emph {et~al.}}]{Giannozzi_2009}%
  \BibitemOpen
  \bibfield  {author} {\bibinfo {author} {\bibfnamefont {Paolo}\ \bibnamefont
  {Giannozzi}} \emph {et~al.},\ }\bibfield  {title} {\enquote {\bibinfo {title}
  {{QUANTUM} {ESPRESSO}: a modular and open-source software project for quantum
  simulations of materials},}\ }\href {\doibase 10.1088/0953-8984/21/39/395502}
  {\bibfield  {journal} {\bibinfo  {journal} {J. Phys. Condens. Matter}\
  }\textbf {\bibinfo {volume} {21}},\ \bibinfo {pages} {395502} (\bibinfo
  {year} {2009})}\BibitemShut {NoStop}%
\bibitem [{\citenamefont {Giannozzi}\ \emph {et~al.}(2017)\citenamefont
  {Giannozzi} \emph {et~al.}}]{Giannozzi_2017}%
  \BibitemOpen
  \bibfield  {author} {\bibinfo {author} {\bibfnamefont {P}~\bibnamefont
  {Giannozzi}} \emph {et~al.},\ }\bibfield  {title} {\enquote {\bibinfo {title}
  {Advanced capabilities for materials modelling with quantum {ESPRESSO}},}\
  }\href {\doibase 10.1088/1361-648x/aa8f79} {\bibfield  {journal} {\bibinfo
  {journal} {J. Phys. Condens. Matter}\ }\textbf {\bibinfo {volume} {29}},\
  \bibinfo {pages} {465901} (\bibinfo {year} {2017})}\BibitemShut {NoStop}%
\bibitem [{\citenamefont {Bl{\"o}chl}(1994)}]{PhysRevB.50.17953PAW1}%
  \BibitemOpen
  \bibfield  {author} {\bibinfo {author} {\bibfnamefont {P.~E.}\ \bibnamefont
  {Bl{\"o}chl}},\ }\bibfield  {title} {\enquote {\bibinfo {title} {Projector
  augmented-wave method},}\ }\href {\doibase 10.1103/PhysRevB.50.17953}
  {\bibfield  {journal} {\bibinfo  {journal} {Phys. Rev. B}\ }\textbf {\bibinfo
  {volume} {50}},\ \bibinfo {pages} {17953--17979} (\bibinfo {year}
  {1994})}\BibitemShut {NoStop}%
\bibitem [{\citenamefont {Kresse}\ and\ \citenamefont
  {Joubert}(1999)}]{PhysRevB.59.1758PAW2}%
  \BibitemOpen
  \bibfield  {author} {\bibinfo {author} {\bibfnamefont {G.}~\bibnamefont
  {Kresse}}\ and\ \bibinfo {author} {\bibfnamefont {D.}~\bibnamefont
  {Joubert}},\ }\bibfield  {title} {\enquote {\bibinfo {title} {From ultrasoft
  pseudopotentials to the projector augmented-wave method},}\ }\href {\doibase
  10.1103/PhysRevB.59.1758} {\bibfield  {journal} {\bibinfo  {journal} {Phys.
  Rev. B}\ }\textbf {\bibinfo {volume} {59}},\ \bibinfo {pages} {1758--1775}
  (\bibinfo {year} {1999})}\BibitemShut {NoStop}%
\bibitem [{\citenamefont {Perdew}\ \emph {et~al.}(1996)\citenamefont {Perdew},
  \citenamefont {Burke},\ and\ \citenamefont {Ernzerhof}}]{GGA-PBE1996}%
  \BibitemOpen
  \bibfield  {author} {\bibinfo {author} {\bibfnamefont {John~P.}\ \bibnamefont
  {Perdew}}, \bibinfo {author} {\bibfnamefont {Kieron}\ \bibnamefont {Burke}},
  \ and\ \bibinfo {author} {\bibfnamefont {Matthias}\ \bibnamefont
  {Ernzerhof}},\ }\bibfield  {title} {\enquote {\bibinfo {title} {Generalized
  gradient approximation made simple},}\ }\href {\doibase
  10.1103/PhysRevLett.77.3865} {\bibfield  {journal} {\bibinfo  {journal}
  {Phys. Rev. Lett.}\ }\textbf {\bibinfo {volume} {77}},\ \bibinfo {pages}
  {3865--3868} (\bibinfo {year} {1996})}\BibitemShut {NoStop}%
\bibitem [{\citenamefont {Slater}\ and\ \citenamefont
  {Koster}(1954)}]{slater-koster}%
  \BibitemOpen
  \bibfield  {author} {\bibinfo {author} {\bibfnamefont {J.~C.}\ \bibnamefont
  {Slater}}\ and\ \bibinfo {author} {\bibfnamefont {G.~F.}\ \bibnamefont
  {Koster}},\ }\bibfield  {title} {\enquote {\bibinfo {title} {Simplified
  {LCAO} method for the periodic potential problem},}\ }\href {\doibase
  10.1103/PhysRev.94.1498} {\bibfield  {journal} {\bibinfo  {journal} {Phys.
  Rev.}\ }\textbf {\bibinfo {volume} {94}},\ \bibinfo {pages} {1498--1524}
  (\bibinfo {year} {1954})}\BibitemShut {NoStop}%
\end{thebibliography}


\end{document}